\documentclass[aps,pre,twocolumn,amsmath,amssymb]{revtex4-2}

\usepackage{mathrsfs}
\usepackage{graphicx}
\usepackage{adjustbox}
\usepackage{longtable}
\usepackage{booktabs}
\usepackage{bm}
\usepackage{threeparttable}
\usepackage{amsmath,amssymb}
\usepackage{color}
\usepackage{epstopdf}
\usepackage{microtype}
\usepackage{tikz}
\usepackage{mathtools}
\usepackage[colorlinks=true,linkcolor=blue,anchorcolor=blue, citecolor=blue,urlcolor=blue]{hyperref}

\def\T{{\theta_T}}
\def\Tp#1{{\theta_T^{#1}}}
\def\h{\Delta t}
\def\n{\lambda}
\def\a{\alpha}
\def\b{\beta}
\def\g{\gamma}

\def\s{\sigma}
\def\th{\theta}
\def\vp{\varphi}

\def\pr{\prime}
\def\ovl#1{\overline{#1}}

\def\df{{\rm d}}

\def\bk#1{\left\langle#1\right\rangle}
\def\ave#1{\mathbb{E}\left[ #1 \right]}
\def\bbkk#1{\kern 0.1em \left\langle \kern -0.3em \left\langle \kern 0.1em #1 \kern 0.1em \right\rangle \kern -0.3em \right\rangle \kern 0.1em}
\def\bko#1{\left( #1 \right)_0}
\def\vb#1{\bm{#1}}
\def\IntV{\int_{-\infty}^{+\infty}d\vb{v}}
\def\pd#1{\partial_{#1}}
\def\pdu#1#2{\partial_{#1}^{#2}} 
\def\pdv#1{\partial_{#1}^{\perp}} 
\def\pdo#1#2{\bko{\pd{#1}{#2}}}
\def\kr#1{\delta_{#1}}
\def\kru#1{\delta^{#1}}
\def\krv#1{\delta_{#1}^{\perp}}
\def\ep#1{\varepsilon_{#1}}

\allowdisplaybreaks[4]
\begin{document}

\preprint{APS/123-QED}

\title{{Mesoscale model of a three-dimensional odd fluid}}

\author{Yuxing Jiao}
\email{jiaoyuxing22@mails.ucas.ac.cn}
\affiliation{Beijing National Laboratory for Condensed Matter Physics and Laboratory of Soft Matter Physics, Institute of Physics, Chinese Academy of Sciences, Beijing 100190, China}\affiliation{School of Physical Sciences, University of Chinese Academy of Sciences, Beijing 100049, China}

\author{Mingcheng Yang}
\email{mcyang@iphy.ac.cn}
\affiliation{Beijing National Laboratory for Condensed Matter Physics and Laboratory of Soft Matter Physics, Institute of Physics, Chinese Academy of Sciences, Beijing 100190, China}\affiliation{School of Physical Sciences, University of Chinese Academy of Sciences, Beijing 100049, China}

\begin{abstract}
Odd fluids are a class of fluids characterized by non-zero antisymmetric transport coefficient tensors induced by broken time-reversal symmetry. In our previous work, a mesoscale simulation model for two-dimensional isotropic odd fluids was developed. Here, we extend the model to the three-dimensional case that corresponds to an anisotropic odd fluid with cylindrical symmetry. Using kinetic theory, we analytically derive the viscosity tensor and Navier-Stokes equation for the three-dimensional mesoscale odd fluid, which are quantitatively verified by simulations. Furthermore, through both simulation and hydrodynamic theory, we demonstrate that the planar Poiseuille flow of the three-dimensional odd fluid exhibits exotic transport behavior. This work thus paves the way for performing large-scale simulations to explore and exploit intriguing phenomena of odd fluids.  
\end{abstract}

\maketitle

\section{Introduction}
Different from normal fluids, odd fluids possess microscopic dynamics with broken time-reversal and parity symmetries. This endows odd fluids with non-zero antisymmetric terms of transport coefficient tensors, according to the Onsager-Casimir reciprocal relations~\cite{NET}, which are odd under time reversal. Examples of such odd fluids include electron Hall fluids~\cite{Avron1995,Hoyos2012,Bandurin2019,Holder2019}, polyatomic gases within a magnetic field~\cite{PolyGas,PolyGas3_Dufor,PolyGas4_thermaldiffusion}, chiral active fluids~\cite{ActiveColloidal,Vitelli_Fluhydro,Markovich2021,Hargus2021}, and so on. The odd transport coefficients can generate fluxes perpendicular to the corresponding non-equilibrium driving forces, thereby enriching hydrodynamics and transportation~\cite{PolyGas,PolyGas3_Dufor,PolyGas4_thermaldiffusion,OddEffects1_HallTransport,OddEffects2_TopoTrans,OddEffects3_TopoWave,OddEffects4_OddMob,OddEffects5_Turb,Banerjee2017-fe,Hargus2021}. Furthermore, this also implies that mixtures of odd fluids with immersed mesoscale objects~\cite{OddEffects2_TopoTrans,Yang_2021} (i.e., odd complex fluids) may exhibit more diverse response and dynamics than conventional complex fluids.

The current simulation studies of odd fluids are primarily confined to molecular-dynamics-type (MD) methods~\cite{OddEffects1_HallTransport,OddEffects2_TopoTrans,Yang_2021,OddEffects4_OddMob,Hargus2021}. Although such simulations can properly describe all microscopic properties of the systems, the cross-scale interactions and detailed evolutionary dynamics make it challenging and even impossible for using  the MD-type approach in large-scale simulation studies of odd fluids and odd complex fluids. However, in studies of fluid dynamics and complex fluid systems, the slow dynamic collective-motion modes (i.e., the hydrodynamic modes) are essential, and the microscopic details of the fluids are unimportant. Motivated by the considerations, the mesoscopic simulation models for the normal fluids have been developed over the past few decades, where real fluids are coarse-grained but the essential hydrodynamic behaviors are retained. Prominent approaches include the lattice Boltzmann method~\cite{LBE,Ladd1_1994,Ladd2_1994}, dissipative particle dynamics~\cite{DPD,Hoogerbrugge_1992}, and multi-particle collision dynamics (MPC)~\cite{MPC1,MPC2,MPC_MD,Kapral2008,Gompper_2009}, each considerably improving the simulation efficiency of the traditional complex fluids. 

Very recently, to address the lack of a coarse-grained simulation approach for two-dimensional  (2D) odd fluids, we developed a mesoscale odd fluid model, named chiral stochastic rotation dynamics (CSRD)~\cite{CSRD}, by extending the stochastic rotation dynamics (SRD), a widely-used version of MPC. In our previous work, we demonstrated that the CSRD correctly captures all the features of 2D odd fluids. However, three-dimensional (3D) odd fluids are more prevalent in the real world and hold greater significance in terms of practical applications. Moreover, unlike their 2D counterparts, symmetry dictates that 3D odd fluids cannot be isotropic~\cite{Avron1998,Vitelli}, thereby allowing more transport coefficients to emerge. Therefore, it is of fundamental importance and interest to develop a 3D mesoscale odd fluid model and explore its intricate transport behaviors. 

In this paper, we propose a 3D mesoscale odd fluid by extending the 2D-CSRD model to 3D case. Through a kinetic theory, we then derive the Navier-Stokes equation and viscosity expressions for the 3D-CSRD model, which are quantitatively verified by performing 3D-CSRD simulations. Furthermore, as a typical case study, we employ the 3D-CSRD method to investigate the planar Poiseuille flow of 3D odd fluids, revealing anomalous transport behaviors that are in excellent agreement with hydrodynamic theory.

\section{3D mesoscale odd fluid: 3D-CSRD model}\label{Sec::TheModel}
In 3D fluids, the existence of nonzero odd viscosities requires the breaking of isotropy. Here, we consider an anisotropic 3D fluid with the cylindrical symmetry (i.e., the rotational symmetry around a fixed axis, say the $z$-axis) and the broken mirror symmetry about planes including $z$-axis, since such odd fluids are the most common and the easiest to realize~\cite{PolyGas,NET,Markovich2021,Vitelli}. The CSRD fluid is a particle-based mesoscale simulation model that consists of a set of $\mathcal{N}$ point particles of mass $m$. The position and velocity of particle $i$ are denoted by $\bm{r}_i$ and $\bm{v}_i$ respectively, and they are updated through two alternate steps: the streaming step and the collision step. In the streaming step, the particles move freely:
\begin{equation}\label{strEq}
    \bm{r}_i\left( t+\h \right)=\bm{r}_i(t)+\bm{v}_i(t)\h,
\end{equation}
with $\h$ the time step. In the collision step, the particles are first sorted into the cells of a cubic lattice sized $l$ according to their positions. As in the traditional SRD, the cubic lattice should be randomly shifted for every CSRD step in order to preserve the Galilean invariance~\cite{MPC_RS}. Then in each cell, the following rotation-type operation is performed on the velocities of the particles:
\begin{equation}\label{colEq}
    \bm{v}_i\left( t+\h \right)=\bm{v}_{cm}+\bm{R}\cdot\left(\bm{v}_i(t)-\bm{v}_{cm}\right),
\end{equation}
where $\bm{v}_{cm}$ is the center-of-mass velocity of the cell, and $\bm{R}$ is a rotation operator. The rotation consists of two parts: $\bm{R}=\bm{R}^{(2)}\cdot\bm{R}^{(1)}$. Here, $\bm{R}^{(1)}=\bm{R}^{(1)}\left( \bm{n},\omega \right)$ refers to a rotation arond a random axis $\bm{n}$, uniformly distributing on the surface of a unit sphere, by a fixed angle $\omega$; while $\bm{R}^{(2)}=\bm{R}^{(2)}\left( \bm{e}_z,\theta \right)$ is an additional rotation around the $z$-axis by an angle $\theta$. 

{  We comment on the physical motivation for the introduction of these two rotation operations. The random rotation $\bm{R}^{(1)}$ follows the conventional SRD scheme, which mimics the velocity changes during real collisions among fluid particles in normal isotropic fluids. However, in odd fluids, the collisions among fluid particles break time-reversal and parity symmetries. To realize such collisions in CSRD, we introduce an additional rotation $\bm{R}^{(2)}$ after the normal isotropic rotation $\bm{R}^{(1)}$. Because we focus on 3D odd fluids with $C_\infty$ cylindrical symmetry, we choose the $z$-axis as the rotation axis, yielding $\bm{R}^{(2)}=\bm{R}^{(2)}\left( \bm{e}_z, \theta \right)$, where $\theta$ characterizes the strength of the chirality.}

The introduction of the additional rotation $\bm{R}^{(2)}$ breaks time-reversal and parity symmetries of the CSRD, rendering the 3D-CSRD fluid is anisotropic with the $C_\infty$ cylindrical symmetry. Consequently, 3D-CSRD may exhibit transport coefficients forbidden in 3D isotropic fluids, for example, the odd viscosities. The CSRD model will reduce to conventional SRD model when $\theta=0$. 

The CSRD inherits all equilibrium properties of the SRD, and satisfies the particle number conservation and cell-level momentum/energy conservation. In the absence of non-equilibrium drivings, the CSRD fluid rapidly relaxes to equilibrium state with the Maxwellian distribution and the ideal-gas equation of state~\cite{CSRD}. However, the symmetry of 3D-CSRD is different form the SRD. Before providing the hydrodynamic equations of the 3D-CSRD, we briefly introduce the stress constitutive relation for the odd fluid with the $C_\infty$ symmetry.

\section{Constitutive relation for odd fluid with \texorpdfstring{$C_\infty$}{C\_infty} symmetry}\label{Sec::CinfFluids}
Generally, the stress of a Newtonian fluid is composed of a hydrostatic part and a viscous part:
\begin{equation}
    \sigma_{\a\b}=\sigma_{\a\b}^{\text{h}}+\sigma_{\a\b}^{\text{v}}.
\end{equation}
The former is the stress nonvanishing even in the fluid without any disturbance; while the latter describes the linear response to velocity gradients $\dot{e}_{\mu\nu}\triangleq\pd{\nu}u_\mu$:
\begin{equation}\label{VisCon}
    \sigma_{\a\b}^{\text{v}}=\eta_{\a\b\mu\nu}\dot{e}_{\mu\nu},
\end{equation}
where $\eta_{\a\b\mu\nu}$ is the viscosity tensor.

By considering the constraint of the fluid's symmetry, the forms of $\sigma_{\a\b}^{\text{h}}$ and $\eta_{\a\b\mu\nu}$ can be simplified. For fluids with the $C_\infty$ symmetry, their hydrostatic tensor and viscous tensor must be invariant under the $C_\infty$-group transformations (say, the rotations about the $z$-axis). The simplified forms of the two tensors has been obtained in a very recent work by Vitelli et al.~\cite{Vitelli}. Here, we briefly summarize their results. 

In general, the rank-2 tensors are represented by nine tensor product bases $\bm{e}_\a\otimes\bm{e}_\b$. For example, a rank-2 tensor $\bm{T}$ is expressed as $\bm{T}=\sum_\a\sum_\b T_{\a\b}\bm{e}_\a\otimes\bm{e}_\b$ and its components can be arranged as a $3\times 3$ matrix. To simplify, we construct a new set of bases $\left\{ \psi^I \right\}$ ($I\in\left\{ 1,2,\dots,9 \right\}$) by the following linear combination of the tensor product bases:
\begin{equation}
    \psi^I=X_{\a\b}^I\left( \bm{e}_\a\otimes\bm{e}_\b \right).
\end{equation} 
Under the new basis $\left\{ \psi^I \right\}$, a rank-2 tensor $\bm{T}$ can be expressed by a $9\times 1$ vector, $\bm{T}=\sum_I T^I\psi^I$. Here, the coefficients $X_{\a\b}^I$ of this linear combination are determined by the irreducible decomposition of the tensor product of two $O(3)$ group's vector representations. Following the results of Vitelli et al.~\cite{Vitelli}, the transformation between these two bases reads:
\begin{equation}\label{BisTran}
    \psi^I=\frac{1}{2}\tau_{\a\b}^I\left( \bm{e}_\a\otimes\bm{e}_\b \right),\qquad\bm{e}_\a\otimes\bm{e}_\b=\tau_{\a\b}^I\psi^I.
\end{equation}
Herein, the coefficients are denoted by the matrices $\tau_{\a\b}^I$—the unnormalized Clebsch–Gordan coefficients:
\begin{equation}
    \begin{aligned}
        &\tau_{\a\b}^1=\sqrt{\frac{2}{3}}\kr{\a\b},\\
        \tau_{\a\b}^2=\ep{x\a\b},\qquad&\tau_{\a\b}^3=\ep{y\a\b},\qquad\tau_{\a\b}^4=\ep{z\a\b},\\
        \,\qquad\bm{\tau}^5=
        \begin{bmatrix}
            1 &  0 & 0 \\
            0 & -1 & 0 \\
            0 &  0 & 0
        \end{bmatrix},&\qquad\quad
        \bm{\tau}^6=
        \begin{bmatrix}
            0 & 1 & 0 \\
            1 & 0 & 0 \\
            0 & 0 & 0
        \end{bmatrix},\\
        \bm{\tau}^7=\frac{-1}{\sqrt{3}}
        \begin{bmatrix}
            1 &  0 & 0 \\
             0 & 1 & 0 \\
             0 &  0 & -2
        \end{bmatrix},\,
        &\bm{\tau}^8=
        \begin{bmatrix}
            0 & 0 & 0 \\
            0 & 0 & 1 \\
            0 & 1 & 0
        \end{bmatrix},\,
        \bm{\tau}^9=
        \begin{bmatrix}
            0 & 0 & 1 \\
            0 & 0 & 0 \\
            1 & 0 & 0
        \end{bmatrix},
    \end{aligned}
\end{equation}
with $\kr{\a\b}$ the Kronecker delta and $\ep{\a\b\g}$ the rank-3 Levi-Civita tensor. $\tau_{\a\b}^I$ holds the following orthogonality relations:
\begin{equation}\label{Orth}
    \tau_{\a\b}^I\tau_{\a\b}^J=2\kru{IJ},\qquad\tau_{\a\b}^I\tau_{\mu\nu}^I=2\kr{\a\mu}\kr{\b\nu},
\end{equation}
which implies the inner product of bases $\psi^I$ and $\psi^J$ is $\bk{\psi^I,\psi^J}=\frac{1}{2}\kru{IJ}$. In the representation of $\psi^I$, the nine components of a rank-2 tensor are classified into three parts: the scalar part represented by $\psi^1$, the pseudovector part represented by $\psi^{2\text{--}4}$, and the symmetric traceless part represented by $\psi^{5\text{--}9}$. With the help of Eqs.~\eqref{BisTran}, the viscous stress constitutive relation Eq.~\eqref{VisCon} under the representation of $\psi^I$ reads:
\begin{equation}\label{ConstitutiveRelation}
    \sigma^{\text{v},I}=\eta^{IJ}\dot{e}^J,
\end{equation}
with
\begin{equation}\label{Tran_ConstitutiveRelation}
    \sigma^{\text{v},I}=\sigma_{\a\b}^{\text{v}}\tau_{\a\b}^I,\quad \dot{e}^I=\dot{e}_{\a\b}\tau_{\a\b}^I,\quad \eta^{IJ}=\frac{1}{2}\tau_{\a\b}^I\eta_{\a\b\mu\nu}\tau_{\mu\nu}^J.
\end{equation}

By using the new set of bases $\left\{ \psi^I \right\}$, we now obtain the general expressions of the $C_\infty$-symmetric hydrostatic stress tensor and viscosity tensor. The simplified hydrostatic stress tensor is:
\begin{equation}\label{HStress}
    \sigma_{\a\b}^{\text{h}}=-P\kr{\a\b}-\tau_z\ep{z\a\b}+\gamma\tau_{\a\b}^7,
\end{equation}
where $P$, $\tau_z$, and $\gamma$ are hydrostatic pressure, torque, and shear stress, respectively. The simplified form of viscosity $\eta^{IJ}$ is:
\begin{widetext}
    \begin{equation}\label{VisT_Psi}
        \bm{\eta}=2 \left[
        \begin{smallmatrix}
            3\zeta/2 & 0 & 0 & \eta_A^e-\eta_A^o & 0 & 0 & \eta_s^e+\eta_s^o & 0 & 0 \\ 
            0 & \eta_{R,1} & \eta_R^o & 0 & 0 & 0 & 0 & \eta_{Q,1}^e+\eta_{Q,1}^o & \eta_{Q,2}^e+\eta_{Q,2}^o \\ 
            0 & -\eta_R^o & \eta_{R,1} & 0 & 0 & 0 & 0 & \eta_{Q,2}^e+\eta_{Q,2}^o & -\eta_{Q,1}^e-\eta_{Q,1}^o \\ 
            \eta_A^e+\eta_A^o & 0 & 0 & \eta_{R,2} & 0 & 0 & \eta_{Q,3}^e+\eta_{Q,3}^o & 0 & 0 \\ 
            0 & 0 & 0 & 0 & \mu_1 & \eta_1^o & 0 & 0 & 0 \\ 
            0 & 0 & 0 & 0 & -\eta_1^o & \mu_1 & 0 & 0 & 0 \\ 
            \eta_s^e-\eta_s^o & 0 & 0 & \eta_{Q,3}^e-\eta_{Q,3}^o & 0 & 0 & \mu_3 & 0 & 0 \\ 
            0 & \eta_{Q,1}^e-\eta_{Q,1}^o & \eta_{Q,2}^e-\eta_{Q,2}^o & 0 & 0 & 0 & 0 & \mu_2 & \eta_2^o \\ 
            0 & \eta_{Q,2}^e-\eta_{Q,2}^o & -\eta_{Q,1}^e+\eta_{Q,1}^o & 0 & 0 & 0 & 0 & -\eta_2^o & \mu_2 \\ 
        \end{smallmatrix}
            \right].
    \end{equation}
\end{widetext}
Note that there are 19 allowed viscosities, and the viscosities labeled by the superscript $o$ are the odd transport coefficients.
{  Based on the transformation properties of these 19 viscosities under mirror reflections about planes containing the $z$-axis, we can classify them into two categories. For convenience, we take the reflection $y\to-y$ as an example. Viscosities that are invariant under this reflection (e.g., $\mu_1(y)=\mu_1(-y)$) are the parity-preserving viscosities, while viscosities that change sign under the reflection (e.g., $\eta_1^o(y)=-\eta_1^o(-y)$) are the parity-violating viscosities. The parity-preserving viscosities include:
\begin{equation}
    \begin{aligned}
        \zeta,\,\eta_{R,1},\,\eta_{R,2},\,\eta_{R,3}&,\,\mu_1,\,\mu_2,\,\mu_3,\,\eta_{Q,1}^e,\,\eta_s^e,\\
        &\eta_{Q,1}^o,\,\eta_s^o.
    \end{aligned}
\end{equation}
The parity-violating viscosities are:
\begin{equation}
    \begin{aligned}
        \eta_A^e,\,&\eta_{Q,2}^e,\,\eta_{Q,3}^e,\\
        \eta_R^o,\,\eta_1^o,\,&\eta_2^o,\,\eta_A^o,\,\eta_{Q,2}^o,\,\eta_{Q,3}^o.
    \end{aligned}
\end{equation}
If the fluid is microscopically reversible, the viscosity tensor in Eq.~(12) satisfies an additional symmetry dictated by the Onsager-Casimir reciprocal relations~\cite{NET}. The microscopic reversibility here means that the microscopic dynamics is invariant under time reversal combined with the inverse of chirality, such as the reverse of enternal magnetic field $\bm{B}$ and the overall rotational angular velocity $\bm{\Omega}$. Without loss of generality, we assume that isotropy is broken by an angular velocity $\bm{\Omega}$ along the $z$-axis. The Onsager-Casimir reciprocal relations require $\eta^{IJ}(\bm{\Omega})={\eta}^{JI}(-\bm{\Omega})$. We note that the reflection $y\to-y$ corresponds to reversing the angular velocity ($\bm{\Omega}\to-\bm{\Omega}$), which implies ${\eta}^{IJ}(y)={\eta}^{JI}(-y)$. Therefore, the parity-preserving viscosities must correspond to the symmetric components of $\eta^{IJ}$, while the parity-violating viscosities correspond to the antisymmetric components. Thus, we conclude that, in microscopically reversible systems, the odd transport coefficients within the parity-preserving class ($\eta_{Q,1}^o,\,\eta_s^o$) and the even transport coefficients within the parity-violating class ($\eta_A^e,\,\eta_{Q,2}^e,\,\eta_{Q,3}^e$) must both vanish.

We finally provide a physical illustration for these nonzero odd viscosities. Physically, the viscosity mainly arises from the momentum transport induced by microscopic collisions between fluid particles. In contrast to normal fluids, collisions among particles in odd fluids with $C_\infty$ cylindrical symmetry are chiral. This chiral scattering can give rise to the transverse momentum transport, characterized by the off-diagonal elements of the viscosity tensor $\eta^{IJ}$, such as the odd viscosities $\eta^o_1$ and $\eta^o_2$. More specifically, for example, $\eta^o_1$ corresponds to the matrix element $\eta^{56}$, which relates the simple shear rate $\dot{e}^6=\dot{e}_{\a\b}\tau_{\a\b}^6=\pd{x}u_y+\pd{y}u_x$ to the pure shear stress $\sigma^{\text{v},5}=\sigma_{\a\b}^{\text{v}}\tau_{\a\b}^5=\sigma_{xx}^{\text{v}}-\sigma_{yy}^{\text{v}}$. The physical interpretations of the remaining viscosities can be deduced in an analogous manner.

\section{Theoretical results}\label{Sec::TheoreticalResults}
The hydrodynamic equations of the 3D-CSRD can be derived from the kinetic theory. Here, we employ a non-equilibrium kinetic approach, which was previously utilized by Pooley and Yeomans to derive the hydrodynamics and transport coefficients of the conventional SRD~\cite{Pooley}. In this method, we assume the system is in a state of local equilibrium, where the single-particle distribution function is characterized by the local density, flow velocity field, and temperature. Under the molecular chaos assumption, we evaluate how the distributions of conserved quantities evolve under the 3D-CSRD dynamics [Eqs.~\eqref{strEq} and \eqref{colEq}]. This allows us to determine the specific forms of the conserved fluxes, from which the macroscopic hydrodynamic equations and transport coefficients are ultimately obtained. In this section, we summarize the main results of this lengthy kinetic derivation, including the continuity equation, the Navier-Stokes equation, and the viscosity tensor of the 3D-CSRD, while leaving the exhaustive derivation details in Appendix~\ref{Appendix::KineticDerivation} for reference.

The hydrodynamic equations (including the continuity equation and the Navier-Stokes equation) for the 3D-CSRD are generally written as
\begin{equation}\label{conservationEq_s}
    \begin{aligned}
        \frac{\mathrm{d}\rho}{\mathrm{d}t} &= -\rho\pd{\alpha} u_\alpha,\\
        \rho\frac{\mathrm{d}u_\alpha}{\mathrm{d}t} &= \pd{\beta}\sigma_{\alpha\beta},
    \end{aligned}
\end{equation}
where $\rho=\rho(\bm{r},t)$ is the mass density field and $u_\alpha=u_\alpha(\bm{r},t)$ denotes the flow velocity field. The stress tensor $\sigma_{\alpha\beta}$ splits into a hydrostatic part and a viscous part: $\sigma_{\a\b}=\sigma_{\a\b}^{\text{h}}+\sigma_{\a\b}^{\text{v}}$. Furthermore, according to the momentum fluxes originating from the streaming step and the collision step, the stress tensor can also be naturally decomposed into a kinetic part (denoted by the superscript $\text{kin}$) and a collisional part (denoted by the superscript $\text{col}$), respectively. This double decomposition can be explicitly written as:
\begin{equation}\label{HydrostaticStress_s}
    \begin{aligned}
        \sigma_{\a\b}&=\sigma_{\a\b}^{\text{h}}+\sigma_{\a\b}^{\text{v}}\\
                     &=\left( \sigma_{\a\b}^{\text{h,kin}} + \sigma_{\a\b}^{\text{h,col}} \right) + \left( \sigma_{\a\b}^{\text{v,kin}} + \sigma_{\a\b}^{\text{v,col}} \right).
    \end{aligned}
\end{equation}

We first illustrate the hydrostatic stress, which is purely a pressure term originating solely from the streaming step:
\begin{equation}\label{HStress_s}
    \sigma_{\a\b}^{\text{h}}=\sigma_{\a\b}^{\text{h,kin}}=-P\kr{\a\b}.
\end{equation}
The pressure here follows the equation of state for an ideal gas:
\begin{equation}\label{Pressure_s}
    P=\rho k_B T/m,
\end{equation}
where $k_B$ is Boltzmann's constant, $T$ is the temperature, and $m$ is the mass of a fluid particle. We remark that the hydrostatic torque and shear stress are absent in the 3D-CSRD because the additional rotation operation $\bm{R}^{(2)}$ does not generate a net torque or shear. This highlights the equilibrium nature of the model and distinguishes it from the chiral active fluids~\cite{Vitelli_Fluhydro}, which possess a nonvanishing hydrostatic torque driven by their activity.

The viscous stress is $\sigma_{\a\b}^{\text{v}}=\sigma_{\a\b}^{\text{v,kin}} + \sigma_{\a\b}^{\text{v,col}}$, where the kinetic and collisional parts are both nonzero and can be expressed as $\sigma_{\a\b}^{\text{v,kin}}=\eta^{\text{kin}}_{\a\b\mu\nu}\dot{e}_{\mu\nu}$ and $\sigma_{\a\b}^{\text{v,col}}=\eta^{\text{col}}_{\a\b\mu\nu}\dot{e}_{\mu\nu}$. Therefore, the viscosity tensor consists of two parts, and the viscous constitutive relation is written as:
\begin{equation}\label{genConR_s}
    \sigma_{\a\b}^{\text{v}}=\eta_{\a\b\mu\nu}\dot{e}_{\mu\nu}=\left( \eta^{\text{kin}}_{\a\b\mu\nu} + \eta^{\text{col}}_{\a\b\mu\nu} \right)\dot{e}_{\mu\nu}.
\end{equation}
Expressed in terms of the basis $\left\{ \psi^I \right\}$ introduced in Section~\ref{Sec::CinfFluids}, the viscosity tensor reads:
\begin{widetext}
    \begin{equation}\label{TotVis_s}
        \bm{\eta}=2 \begin{bmatrix}
            3\zeta/2 & 0 & 0 & -\eta_A^o & 0 & 0 & \eta_s^e & 0 & 0 \\ 
            0 & \eta_{R,1} & \eta_R^o & 0 & 0 & 0 & 0 & \eta_{Q,1}^e & \eta_{Q,2}^o \\ 
            0 & -\eta_R^o & \eta_{R,1} & 0 & 0 & 0 & 0 & \eta_{Q,2}^o & -\eta_{Q,1}^e \\ 
            \eta_A^o & 0 & 0 & \eta_{R,2} & 0 & 0 & \eta_{Q,3}^o & 0 & 0 \\ 
            0 & 0 & 0 & 0 & \mu_1 & \eta_1^o & 0 & 0 & 0 \\ 
            0 & 0 & 0 & 0 & -\eta_1^o & \mu_1 & 0 & 0 & 0 \\ 
            \eta_s^e & 0 & 0 & -\eta_{Q,3}^o & 0 & 0 & \mu_3 & 0 & 0 \\ 
            0 & \eta_{Q,1}^e & -\eta_{Q,2}^o & 0 & 0 & 0 & 0 & \mu_2 & \eta_2^o \\ 
            0 & -\eta_{Q,2}^o & -\eta_{Q,1}^e & 0 & 0 & 0 & 0 & -\eta_2^o & \mu_2 \\ 
        \end{bmatrix},
    \end{equation}
\end{widetext}
which shows that there are $14$ nonvanishing viscosities in the total viscosity tensor of the 3D-CSRD. Compared to the general form of the viscosity tensor in Eq.~\eqref{VisT_Psi}, $5$ viscosities ($\eta^e_A$, $\eta^o_s$, $\eta^o_{Q,1}$, $\eta^e_{Q,2}$, and $\eta^e_{Q,3}$) are absent in the 3D-CSRD. As stated in Section~\ref{Sec::CinfFluids}, the vanishing of these viscosities is fully consistent with the microscopic reversibility of CSRD.

The kinetic viscosity tensor contributes to five independent viscosity coefficients, and its matrix representation is given by:
\begin{widetext}
\begin{equation}\label{KinVisM_s}
    \bm{\eta}^{\rm{kin}}=2
    \begin{bmatrix}
        \quad 0 \quad& \quad 0 \quad & \quad 0 \quad & \quad 0 \quad & \quad 0 \quad & \quad 0 \quad & \quad 0 \quad & \quad 0 \quad & \quad 0 \quad \\ 
        0 & 0 & 0 & 0 & 0 & 0 & 0 & 0 & 0 \\ 
        0 & 0 & 0 & 0 & 0 & 0 & 0 & 0 & 0 \\ 
        0 & 0 & 0 & 0 & 0 & 0 & 0 & 0 & 0 \\ 
        0 & 0 & 0 & 0 & \mu_1^{\rm{kin}} & \eta_1^{o,\rm{kin}} & 0 & 0 & 0 \\ 
        0 & 0 & 0 & 0 & -\eta_1^{o,\rm{kin}} & \mu_1^{\rm{kin}} & 0 & 0 & 0 \\ 
        0 & 0 & 0 & 0 & 0 & 0 & \mu_3^{\rm{kin}} & 0 & 0 \\ 
        0 & 0 & 0 & 0 & 0 & 0 & 0 & \mu_2^{\rm{kin}} & \eta_2^{o,\rm{kin}} \\ 
        0 & 0 & 0 & 0 & 0 & 0 & 0 & -\eta_2^{o,\rm{kin}} & \mu_2^{\rm{kin}} \\ 
    \end{bmatrix}.
\end{equation}
The explicit expressions for these five kinetic viscosities separately are:
\begin{equation}\label{KinVis_s}
    \begin{aligned}
            \mu_1^{\rm{kin}}&=\rho\frac{k_BT}{m}\h\left\{ \frac{\n\left( 1-q\cos2\th \right)}{\left( \n-1+e^{-\n} \right)\left[ \left( 1-q\cos2\th \right)^2+q^2\sin^22\th \right]} -\frac{1}{2}\right\},\\
            \mu_2^{\rm{kin}}&=\rho\frac{k_BT}{m}\h\left\{ \frac{\n\left( 1-q\cos\th \right)}{\left( \n-1+e^{-\n} \right)\left[ \left( 1-q\cos\th \right)^2+q^2\sin^2\th \right]} -\frac{1}{2}\right\},\\
            \mu_3^{\rm{kin}}&=\rho\frac{k_BT}{2m}\h\left[ \frac{5\n}{\left( \n-1+e^{-\n} \right)\left( 2-\cos\omega-\cos2\omega \right)} -1\right],\\
            \eta_1^{o,\rm{kin}}&=-\rho\frac{k_BT}{m}\h\frac{\n q\sin2\th}{\left( \n-1+e^{-\n} \right)\left[ \left( 1-q\cos2\th \right)^2+q^2\sin^22\th \right]},\\
            \eta_2^{o,\rm{kin}}&=\rho\frac{k_BT}{m}\h\frac{\n q\sin\th}{\left( \n-1+e^{-\n} \right)\left[ \left( 1-q\cos\th \right)^2+q^2\sin^2\th \right]},
    \end{aligned}
\end{equation}
where $\lambda=\rho l^3/m$ ($l$ being the size of the collision cell) represents the average number of particles per cell, and $q$ is a function of $\omega$ defined as:
\begin{equation}\label{qInKinVis_s}
    q=\frac{1}{5}\left( 1+2\cos\omega+2\cos 2\omega \right).
\end{equation}

The collisional viscosity tensor contributes to all $14$ viscosity coefficients:
\begin{equation}\label{ColVisM_s}
    \bm{\eta}^{\rm{col}}=2 \begin{bmatrix}
        3\zeta^{\rm{col}}/2 & 0 & 0 & -\eta_A^{\rm{o,col}} & 0 & 0 & \eta_s^{\rm{e,col}} & 0 & 0 \\ 
        0 & \eta_{R,1}^{\rm{col}} & \eta_R^{\rm{o,col}} & 0 & 0 & 0 & 0 & \eta_{Q,1}^{\rm{e,col}} & \eta_{Q,2}^{\rm{o,col}} \\ 
        0 & -\eta_R^{\rm{o,col}} & \eta_{R,1}^{\rm{col}} & 0 & 0 & 0 & 0 & \eta_{Q,2}^{\rm{o,col}} & -\eta_{Q,1}^{\rm{e,col}} \\ 
        \eta_A^{\rm{o,col}} & 0 & 0 & \eta_{R,2}^{\rm{col}} & 0 & 0 & \eta_{Q,3}^{\rm{o,col}} & 0 & 0 \\ 
        0 & 0 & 0 & 0 & \mu_1^{\rm{col}} & \eta_1^{\rm{o,col}} & 0 & 0 & 0 \\ 
        0 & 0 & 0 & 0 & -\eta_1^{\rm{o,col}} & \mu_1^{\rm{col}} & 0 & 0 & 0 \\ 
        \eta_s^{\rm{e,col}} & 0 & 0 & -\eta_{Q,3}^{\rm{o,col}} & 0 & 0 & \mu_3^{\rm{col}} & 0 & 0 \\ 
        0 & \eta_{Q,1}^{\rm{e,col}} & -\eta_{Q,2}^{\rm{o,col}} & 0 & 0 & 0 & 0 & \mu_2^{\rm{col}} & \eta_2^{\rm{o,col}} \\ 
        0 & -\eta_{Q,2}^{\rm{o,col}} & -\eta_{Q,1}^{\rm{e,col}} & 0 & 0 & 0 & 0 & -\eta_2^{\rm{o,col}} & \mu_2^{\rm{col}} \\ 
    \end{bmatrix}.
\end{equation}
All these collisional viscosities are linear combinations of the same three factors $\eta_1^{\rm{col}}$, $\eta_2^{\rm{col}}$, and $\eta_3^{\rm{col}}$:
\begin{equation}\label{ColVis_s}
    \begin{aligned}
        &\zeta^{\rm{col}}=\frac{1}{3}\eta_1^{\rm{col}},\\
        &\eta_{R,1}^{\rm{col}}=\mu_2^{\rm{col}}=\frac{1}{2}\eta_1^{\rm{col}}+\frac{1}{4\sqrt{3}}\eta_2^{\rm{col}},\\
        &\eta_{R,2}^{\rm{col}}=\mu_1^{\rm{col}}=\frac{1}{2}\eta_1^{\rm{col}}-\frac{1}{2\sqrt{3}}\eta_2^{\rm{col}},\\
        &\mu_3^{\rm{col}}=\frac{1}{2}\eta_1^{\rm{col}}+\frac{1}{2\sqrt{3}}\eta_2^{\rm{col}},\\
        &\eta_{s}^{e,\rm{col}}=-\frac{2\sqrt{2}}{3}\eta_{Q,1}^{e,\rm{col}}=\frac{1}{\sqrt{6}}\eta_2^{\rm{col}},\\
        &\eta_{1}^{o,\rm{col}}=-2\eta_{2}^{o,\rm{col}}=2\eta_{R}^{o,\rm{col}}=-2\eta_{Q,2}^{o,\rm{col}}=\frac{1}{2}\eta_3^{\rm{col}},\\
        &\eta_A^{o,\rm{col}}=-\sqrt{2}\eta_{Q,3}^{o,\rm{col}}=\frac{1}{\sqrt{6}}\eta_3^{\rm{col}},
    \end{aligned}
\end{equation}
with
\begin{equation}\label{etasInColVis_s}
    \begin{aligned}
        &\eta_1^{\rm{col}}=\frac{m\left( \n-1+e^{-\n} \right)}{108l\h}\left[ 9 - \left( 1+2\cos\omega \right)\left( 1+2\cos\theta \right) \right],\\
        &\eta_2^{\rm{col}}=-\frac{\sqrt{3}m\left( \n-1+e^{-\n} \right)}{108l\h}\left( 1+2\cos\omega \right)\left( 1-\cos\theta \right),\\
        &\eta_3^{\rm{col}}=\frac{m\left( \n-1+e^{-\n} \right)}{36l\h}\left( 1+2\cos\omega \right)\sin\theta.
    \end{aligned}
\end{equation}

Using Eqs.~\eqref{KinVisM_s}, \eqref{KinVis_s}, \eqref{qInKinVis_s}, \eqref{ColVisM_s}, \eqref{ColVis_s}, and \eqref{etasInColVis_s}, one can obtain the complete experssions for every viscosities in the 3D-CSRD, which are summarized in the appendix \ref{Appendix}. 

\end{widetext}

\subsection{Navier-Stokes equation}
After transforming the viscosity tensor~\eqref{TotVis_s} into its representation in the tensor product basis $\left\{ \bm{e}_\a\otimes\bm{e}_\b \right\}$, the constitutive relation in Eq.~\eqref{genConR_s} becomes:
\begin{equation}\label{ConstitutiveRelationByNB_s}
    \begin{aligned}
        \sigma_{\a\b}^{\rm{v}}=&\,\zeta\kr{\a\b}\bm{\nabla}\cdot\bm{u}+\mu_1\left( \pdv{\b}u_\a^\perp+\pdv{\a}u_\b^\perp-\krv{\a\b}\bm{\nabla}_\perp\cdot\bm{u}^\perp \right)\\
        &+\mu_2\left[ \kr{z\b}\left( \pd{z}u_\a^\perp+\pd{\a}u_z^\perp \right)+\kr{z\a}\left( \pd{z}u_\b^\perp+\pd{\b}u_z^\perp \right) \right]\\
            &+\mu_3\left( \krv{\a\b}\bm{\nabla}_\perp\cdot\bm{u}^\perp+2\kr{z\a}\kr{z\b}\pd{z}u_z-\frac{2}{3}\kr{\a\b}\bm{\nabla}\cdot\bm{u} \right)\\
            &+\eta_{R,1}\left[ \pd{\b}u_\a-\pd{\a}u_\b-\left( \pdv{\b}u_\a^\perp-\pdv{\a}u_\b^\perp \right) \right]\\
            &+\eta_{R,2}\left( \pdv{\b}u_\a^\perp-\pdv{\a}u_\b^\perp \right)\\
            &+2\eta_{Q,1}^e\left( \kr{z\b}\pd{z}u_\a^\perp-\kr{z\a}\pdv{\b}u_z \right)\\
            &+\sqrt{2}\eta_s^e\left[ \frac{4}{3}\kr{\a\b}-\left( \kr{\a\b}\bm{\nabla}_\perp\cdot\bm{u}^\perp+\krv{\a\b}\bm{\nabla}\cdot\bm{u} \right) \right]\\
            &+\eta_1^o\left( \pdu{\b}{\ast}u_\a^\perp +\pdv{\b}u_\a^\ast\right)+2\eta_{Q,2}^o\left( \pdu{\b}{\ast}u_\a-\pd{\b}u_\a^\ast\right)\\
            &-\eta_2^o\left[ \kr{z\a}\left( \pdu{\b}{\ast}u_z+\pd{z}u_\b^\ast \right)+\kr{z\b}\left( \pdu{\a}{\ast}u_z+\pd{z}u_\a^\ast \right) \right]\\
            &+\eta_R^o\left( \pd{\b}u_\a^\ast+\pdu{\b}{\ast}u_\a-\pd{\a}u_\b^\ast+\pdu{\a}{\ast}u_\b \right),
    \end{aligned}
\end{equation}
where we have introduced the notations $\delta^{\perp}_{\a\b}=\delta_{\a\b}-\kr{z\a}\kr{z\b}$, $u_\a^\ast=\ep{z\a\b}u_\b$, $u_\a^{\perp}= \krv{\a\b}u_\b$, $\pdu{\a}{\ast}=\ep{z\a\b}\pd{\b}$, and $\pdv{\a}= \krv{\a\b}\pd{\b}$. We also utilize the relations $\eta_{Q,3}^o=2\eta_{Q,2}^o/\sqrt{3}$ and $\eta_A^o=-\frac{4}{\sqrt{6}}\eta_{Q,2}^o$ implied by Eq.~\eqref{ColVis_s} to simplify the expression by absorbing the terms involving $\eta_{Q,3}^o$ and $\eta_A^o$ into those with $\eta_{Q,2}^o$. Substituting the constitutive relation in Eq.~\eqref{ConstitutiveRelationByNB_s} and the hydrostatic stress in Eq.~\eqref{HydrostaticStress_s} into Eq.~\eqref{conservationEq_s} yields the specific form of the hydrodynamic equations for the 3D-CSRD:
\begin{equation}\label{NSEq_s}
    \begin{aligned}
        \frac{\mathrm{d}\rho}{\mathrm{d}t} =&\, -\rho\bm{\nabla}\cdot\bm{u},\\
        \rho\frac{\mathrm{d}\bm{u}}{\mathrm{d}t}=&\,-\bm{\nabla} P+\hat{\eta}_b\bm{\nabla}\left( \bm{\nabla}\cdot\bm{u} \right)+\hat{\eta}_{zb}\left( \bm{\nabla}\pd{z}u_z+\hat{\bm{e}}_z\pd{z}\bm{\nabla}\cdot\bm{u}\right)\\
        &+\hat{\eta}\bm{\nabla}^2\bm{u}+\hat{\eta}_{zs1}\hat{\bm{e}}_z\bm{\nabla}^2u_z+\hat{\eta}_{zs2}\pdu{z}{2}\bm{u}+\hat{\eta}_{zs3}\hat{\bm{e}}_z\pdu{z}{2}u_z\\
        &+\hat{\eta}_{o}\bm{\nabla}^2\bm{u}^\ast+\hat{\eta}_{ob}\left[ \bm{\nabla}\left( \bm{\nabla}\cdot\bm{u}^\ast \right)+\bm{\nabla}^\ast\left( \bm{\nabla}\cdot\bm{u} \right) \right]\\
        &+\hat{\eta}_{zo}\left( \pdu{z}{2}\bm{u}^\ast+\hat{\bm{e}}_z\pd{z}\bm{\nabla}\cdot\bm{u}^\ast+\bm{\nabla}^\ast\pd{z}u_z \right),
    \end{aligned}
\end{equation}
where the coefficients are defined as:
\begin{align*}
    \hat{\eta}&=\mu_1+\eta_{R,2},\\
    \hat{\eta}_b&=\zeta+\frac{1}{3}\mu_3-\eta_{R,2}-\frac{2\sqrt{2}}{3}\eta_s^e,\\
    \hat{\eta}_{zs1}&=\mu_2-\mu_1+\eta_{R,1}-\eta_{R,2}-2\eta_{Q,1}^e,\\
    \hat{\eta}_{zs2}&=\mu_2-\mu_1+\eta_{R,1}-\eta_{R,2}+2\eta_{Q,1}^e,\\
    \hat{\eta}_{zs3}&=\mu_1+2\mu_3-4\mu_2+\eta_{R,2}-\eta_{R,1},\\
    \hat{\eta}_{zb}&=\mu_2-\mu_3+\eta_{R,2}-\eta_{R,1}+\sqrt{2}\eta_{s}^e,\\
    \hat{\eta}_o&=\frac{1}{2}\eta_1^o+\eta_R^o-2\eta_{Q,2}^o,\\
    \hat{\eta}_{ob}&=\frac{1}{2}\eta_1^o-\eta_R^o,\\
    \hat{\eta}_{zo}&=-\left( \frac{1}{2}\eta_1^o+\eta_2^o \right).
\end{align*}

We remark that terms with viscosities $\left\{ \eta_{R,1},\,\eta_{R,2},\,\eta_{R,3},\,\eta_{Q,1}^e,\,\eta_R^o,\,\eta_A^o,\,\eta_{Q,2}^o,\,\eta_{Q,3}^o \right\}$ in the constitutive relation in Eq.~\eqref{ConstitutiveRelationByNB_s} correspond to the antisymmetric part of the viscous stress tensor. This reflects the fact that angular momentum is not conserved in the 3D-CSRD model, a characteristic that is also shared by the original SRD~\cite{Gompper_2009,AMinSRD}. Although angular momentum is usually not conserved in a wide range of odd fluids—such as chiral active fluids driven by active torques or charged particles experiencing the Lorentz force in a magnetic field—rendering these rotational viscosities physically permissible, we emphasize that care should be taken when the 3D-CSRD model is applied to the study of odd fluids with strict angular momentum  conservation. On the one hand, the existence of these viscosities does not alter the structural form of the Navier-Stokes equation. On the other hand, the flow field may still be quantitatively influenced when the stress tensor is incorporated into the boundary conditions, as previously demonstrated in the original SRD framework~\cite{AMinSRD}.

\subsection{Simplified Navier-Stokes equation}
We can significantly simplify the Navier-Stokes equation in~\eqref{NSEq_s} by setting $\omega=2\pi/3$ and considering the limit $|\theta|\ll 1$. 

First, we notice that $\omega=2\pi/3$ causes the two factors $\eta_2^{\rm{col}}$ and $\eta_3^{\rm{col}}$ in the collisional viscosities in Eq.~\eqref{ColVis_s} to vanish (see Eq.~\eqref{etasInColVis_s}). The physical reason behind this is that the first moment of the rotation matrix is $\mathbb{E}\left[ R_{\a\b} \right]=\left( 1+2\cos\omega \right)\kr{\a\b}$, which becomes exactly zero at $\omega=2\pi/3$. This implies that, in sense of the first moment, the isotropy of the collision step is fully restored, since $\eta^{\mathrm{col}}_{\alpha\beta\mu\nu}\propto \left( \delta_{\alpha\mu} - \mathbb{E}\left[ R_{\alpha\mu} \right] \right)\delta_{\beta\nu}$ (see Eq.~\eqref{colvisTensor}). Consequently, under the molecular chaos hypothesis, the off-diagonal elements of $\bm{\eta}^{\rm{col}}$ in Eq.~\eqref{ColVisM_s} vanish, and the diagonal elements reduce to:
\begin{equation}
    \frac{3}{2}\zeta^{\rm{col}}=\mu_1^{\rm{col}}=\mu_2^{\rm{col}}=\mu_3^{\rm{col}}=\eta_{R,1}^{\rm{col}}=\eta_{R,2}^{\rm{col}}=\frac{1}{2}\eta_1^{\rm{col}}.
\end{equation}

Second, in the limit $|\theta|\ll 1$, one can expand the kinetic viscosities in Eq.~\eqref{KinVis_s} up to first order in $\theta$, $\mathcal{O}\left( \theta \right)$, which yields a remarkable simplification:
\begin{equation}
    \mu_1^{\rm{kin}}=\mu_2^{\rm{kin}}=\mu_3^{\rm{kin}},\qquad \eta_1^{o,\rm{kin}}=-2\eta_2^{o,\rm{kin}}.
\end{equation}

Finally, we introduce the following quantities
\begin{equation}
    \begin{aligned}
        &\mu\triangleq\mu_1=\mu_2=\mu_3,\qquad \eta_R\triangleq\eta_{R,1}=\eta_{R,2},\\
        &\eta_o\triangleq\eta_1^{o}=-2\eta_2^{o},
    \end{aligned}
\end{equation}
to explicitly write the total viscosity tensor as:
\begin{equation}
    \begin{aligned}
        &\eta_{\a\b\mu\nu}\\
        =&\,\zeta\kr{\a\b}\kr{\mu\nu}+\eta_R\left( \kr{\a\mu}\kr{\b\nu}-\kr{\a\nu}\kr{\b\mu} \right)\\
        &+\mu\left( \kr{\a\mu}\kr{\b\nu}+\kr{\a\nu}\kr{\b\mu}-\frac{2}{3}\kr{\a\b}\kr{\mu\nu} \right)\\
        &+\frac{1}{2}\eta_o\left( \ep{z\a\mu}\kr{\b\nu}+\ep{z\a\nu}\kr{\b\mu}+\ep{z\b\mu}\kr{\a\nu}+\ep{z\b\nu}\kr{\a\mu} \right).
    \end{aligned}
\end{equation}
Additionally, we remark that the limit $\lim_{\theta\to 0}\left( -\frac{\eta_1^o}{\eta_2^o} \right)=2$ is fully consistent with the behavior of polyatomic gases in a weak magnetic field~\cite{PolyGas}.

Under these conditions, the viscous stress constitutive relation reduces to:
\begin{equation}
    \begin{aligned}
        \sigma_{\a\b}=&\ \zeta\kr{\a\b}\bm{\nabla}\cdot\bm{u} + \eta_R\left( \pd{\b}u_\a-\pd{\a}u_\b \right)\\
                      &\ +\mu\left( \pd{\b}u_\a+\pd{\a}u_\b - \frac{2}{3}\kr{\a\b}\bm{\nabla}\cdot\bm{u} \right)\\
                      &\ +\frac{1}{2}\eta_o\left( \pd{\b}u_\a^\ast + \pdu{\a}{\ast}u_\b + \pd{\a}u_\b^\ast + \pdu{\b}{\ast}u_\a \right).
    \end{aligned}
\end{equation}
The hydrodynamic equations can now be simplified to:
\begin{equation}
    \begin{aligned}
        \frac{\mathrm{d}\rho}{\mathrm{d}t} &= -\rho\bm{\nabla}\cdot\bm{u},\\
        \rho\frac{\mathrm{d}\bm{u}}{\mathrm{d}t}=&\,-\bm{\nabla} P+\hat{\eta}\bm{\nabla}^2\bm{u}+\hat{\eta}_b\bm{\nabla}\left( \bm{\nabla}\cdot\bm{u} \right)\\
        &+\hat{\eta}_o\left[ \bm{\nabla}^2\bm{u}^\ast+\bm{\nabla}\left( \bm{\nabla}\cdot\bm{u}^\ast \right)+\bm{\nabla}^\ast\left( \bm{\nabla}\cdot\bm{u} \right) \right],
    \end{aligned}
\end{equation}
where the effective transport coefficients are defined as:
\begin{equation}
    \hat{\eta}=\eta_1^{\rm{col}}+\mu^{\rm{kin}},\qquad\hat{\eta}_b=\frac{1}{3}\mu^{\rm{kin}},\qquad\hat{\eta}_o=\frac{1}{2}\eta_1^{o,\rm{kin}}.
\end{equation}
}

\section{Simulation measurement of the viscosities}\label{Sec::Measurement}
In this section, we measure all of the elements of the $C_{\infty}$ viscosity tensor (Eq.~\eqref{VisT_Psi}) in the 3D-CSRD model and compare them with the theoretical results derived from the above kinetic method. We use the non-equilibrium route to calculate these viscosities in simulations. Broadly speaking, we first generate some velocity gradients and quantify the induced stress and then use the constitutive relation Eq.~\eqref{ConstitutiveRelation} to obtain the viscosities.

\subsection{Determination of the viscosities}
We divide the constitutive relation into three parts named Part-$\left( 5, 6 \right)$, Part-$\left( 2, 3, 8, 9 \right)$, and Part-$\left( 1, 4 ,7 \right)$, and measure the viscosities belong to these three parts, respectively. 

Part-$\left( 5, 6 \right)$ is defined as
\begin{equation}\label{Eq56}
    \begin{aligned}
            \left[ \begin{smallmatrix}
                {\s^{\rm{v}}}^5\\
                {\s^{\rm{v}}}^6
            \end{smallmatrix} \right]
            =2\left[ \begin{smallmatrix}
                \mu_1 & \eta_1^o\\
                -\eta_1^o & \mu_1
            \end{smallmatrix} \right]
            \left[ \begin{smallmatrix}
                \pd{x}u_x-\pd{y}u_y\\
                \pd{x}u_y+\pd{y}u_x
            \end{smallmatrix} \right]
    \end{aligned}
\end{equation}
with ${\s^{\rm{v}}}^5={\s^{\rm{v}}}_{xx}-{\s^{\rm{v}}}_{yy}$ and ${\s^{\rm{v}}}^6={\s^{\rm{v}}}_{xy}+{\s^{\rm{v}}}_{yx}$. To determine viscosities $\mu_1$ and $\eta_1^o$, we here impose a velocity gradient $\pd{y}u_x\equiv\gamma_{yx}$ in the corresponding simulation.

Part-$\left( 2, 3, 8, 9 \right)$ takes the form
\begin{equation}\label{Eq2389}
    \begin{aligned}
            \left[ \begin{smallmatrix}
                {\s^{\rm{v}}}^2\\
                {\s^{\rm{v}}}^3\\
                {\s^{\rm{v}}}^8\\
                {\s^{\rm{v}}}^9
            \end{smallmatrix} \right]
            =
            2\left[ \begin{smallmatrix}
                \eta_{R,1}   & \eta_R^o      & \eta_{Q,1}^+ & \eta_{Q,2}^+ \\
                -\eta_R^o    & \eta_{R,1}    & \eta_{Q,2}^+ & -\eta_{Q,1}^+ \\
                \eta_{Q,1}^- & \eta_{Q,2}^-  & \mu_2        & \eta_2^o \\
                \eta_{Q,2}^- & -\eta_{Q,1}^- & -\eta_o^2    & \mu_2
            \end{smallmatrix} \right]
            \left[ \begin{smallmatrix}
                \pd{z}u_y-\pd{y}u_z\\
                \pd{x}u_z-\pd{z}u_x\\
                \pd{z}u_y+\pd{y}u_z\\
                \pd{x}u_z+\pd{z}u_x
            \end{smallmatrix} \right],
    \end{aligned}
\end{equation}
where we have ${\s^{\rm{v}}}^2={\s^{\rm{v}}}_{yz}-{\s^{\rm{v}}}_{zy}$, ${\s^{\rm{v}}}^3={\s^{\rm{v}}}_{zx}-{\s^{\rm{v}}}_{xz}$, ${\s^{\rm{v}}}^8={\s^{\rm{v}}}_{yz}+{\s^{\rm{v}}}_{zy}$, and ${\s^{\rm{v}}}^9={\s^{\rm{v}}}_{xz}+{\s^{\rm{v}}}_{zx}$. We also define $\eta_{Q,1}^+\triangleq\eta_{Q,1}^e+\eta_{Q,1}^o$ and $\eta_{Q,1}^-\triangleq\eta_{Q,1}^e-\eta_{Q,1}^o$. These symbols are applied to other viscosities in the same way. The viscosities in this part are obtained by two independent simulations in which the velocity gradients $\pd{y}u_z\equiv\gamma_{yz}$ and $\pd{z}u_y\equiv\gamma_{zy}$ are imposed, respectively.

Part-$\left( 1, 4, 7 \right)$ is given by
\begin{equation}\label{Eq147}
    \begin{aligned}
            \left[ \begin{smallmatrix}
                {\s^{\rm{v}}}^1\\
                {\s^{\rm{v}}}^4\\
                {\s^{\rm{v}}}^7
            \end{smallmatrix} \right]
            =
            2\left[ \begin{smallmatrix}
                \frac{3}{2}\zeta & \eta_A^-     & \eta_S^+ \\
                \eta_A^+         & \eta_{R,2}   & \eta_{Q,3}^+ \\
                \eta_S^-         & \eta_{Q,3}^- & \mu_3
            \end{smallmatrix} \right]
            \left[ \begin{smallmatrix}
                \sqrt{\frac{2}{3}}\pd{\a}u_\a\\
                \pd{y}u_x-\pd{x}u_y\\
                \frac{1}{\sqrt{3}}\left( 2\pd{z}u_z-\pd{x}u_x-\pd{y}u_y \right)
            \end{smallmatrix} \right],
    \end{aligned}
\end{equation}
where we have ${\s^{\rm{v}}}^1=\sqrt{\frac{2}{3}}\left( {\s^{\rm{v}}}_{xx}+{\s^{\rm{v}}}_{yy}+{\s^{\rm{v}}}_{zz} \right)$, ${\s^{\rm{v}}}^4={\s^{\rm{v}}}_{xy}-{\s^{\rm{v}}}_{yx}$, and ${\s^{\rm{v}}}^7=\frac{1}{\sqrt{3}}\left( 2{\s^{\rm{v}}}_{zz}-{\s^{\rm{v}}}_{xx}-{\s^{\rm{v}}}_{yy} \right)$. Here, we perform three independent simulations to determine the nine viscosities. In the first simulation, we impose the gradient $\pd{y}u_x$ as in the Part-$\left( 5, 6 \right)$. In the second simulation, we impose a pure shear by setting $\pd{z}u_z=-\pd{y}u_y\equiv\alpha$. In the third simulation, we impose a volume deformation with the deformation rate $3\alpha$ by setting $\pd{x}u_x=\pd{y}u_y=\pd{z}u_z\equiv\alpha$.

\subsection{Simulation details}
We nondimensionalize the physical quantities by setting $m=1$, $l=1$, and $k_BT=1$ in the simulations. The viscosities are measured at varying additional rotation angles $\theta$ and fixed parameters $\h=0.1$, $\lambda=10$, $\omega=\pi/3$. The simulations are carried out in a cubic box of size $L_\star=20$.

\subsubsection{Generation of the velocity gradients}
The velocity gradients are generated by applying the Lees-Edwards boundary condition \cite{Evans_Morriss_2008}. The value of the gradients mentioned above is set by $\g_{yx}=\g_{yz}=\g_{zy}=0.003$ and $\alpha=-0.0015$. If the velocity gradient involves a system deformation, such as the pure shear and the volume deformation used in determining the viscosities of Part-$\left( 1, 4, 7 \right)$, the size of the simulation box is changed meanwhile with the corresponding deformation rate. For example, the deformation along $z$-axis is given by
\begin{equation}
    L_z\left( t+\h \right)=\left( 1+\pd{z}u_z \h \right)L_z\left( t \right),
\end{equation}
where $L_z\left( z \right)$ is the size along $z$-axis at time $t$.

Special considerations should be incorporated into the simulation of volume deformation. In the simulation, the particle number is initialized as $\mathcal{N}=\lambda {L_\star}^3$. Considering that the deformation is contractive ($\alpha<0$), we set the initial box volume by $V_0=\left( L_\star+1 \right)^3$. We perform our measurement only when the volume is in a vicinity of $L_\star^3$, i.e., $\left[ 0.95L_\star^3,1.05L_\star^3 \right]$. The contraction will increase the temperature so that we must keep the temperature of the system fixed ($k_BT=1$). Here, we apply the Maxwell-Boltzmann scaling thermostat, a thermostat widely used in the traditional SRD simulations~\cite{MBS_Huang}, in the simulation to realize an isothermal measurement.

\subsubsection{Measurement of the stress}
We measure the stress in the simulations by counting the momentum across a given plane. The method of counting is different between the kinetic part and the collisional part of the stress. The kinetic stress at time $t$, denoted by $\sigma_{\a\b}^{\rm{kin}}(t)$, is calculated by accumulating the net momentum across the $\beta$-plane during a streaming step:
\begin{equation}
    \sigma_{\a\b}^{\rm{kin}}(t)=\frac{m}{\h A_\b}\sum_i \chi_ic_{i,\a}.
\end{equation}
Here, the summation runs for the particles across the $\beta$-plane (a plane with normal voctor along $\beta$-axis) within a step, $A_\b$ is the area of the plane, $\bm{c}\triangleq\bm{v}-\bm{u}$ is the peculiar velocity, and $\chi_i=1$ (or $-1$) if the particle $i$ moves along the same (or opposite) direction of the plane's normal vector. The collisional stress is calculated in a collision cell. If a given $\beta$-plane divides a cell, we record the total change in momentum within the half of the cell during a collision step, thus the collisional stress is computed as,
\begin{equation}
    \sigma_{\a\b}^{\rm{col}}(t)=\frac{m}{l^2\h}\sum_i\left( v_{i,\a}\left( t+\h \right)- v_{i,\a}\left( t \right)\right),
\end{equation}
where the summation runs for every particle in the half of the cell. Then both parts of the stress are averaged over time and across the ensemble.

\begin{figure}[htbp]
    \centering
    \includegraphics[keepaspectratio, width=\columnwidth]{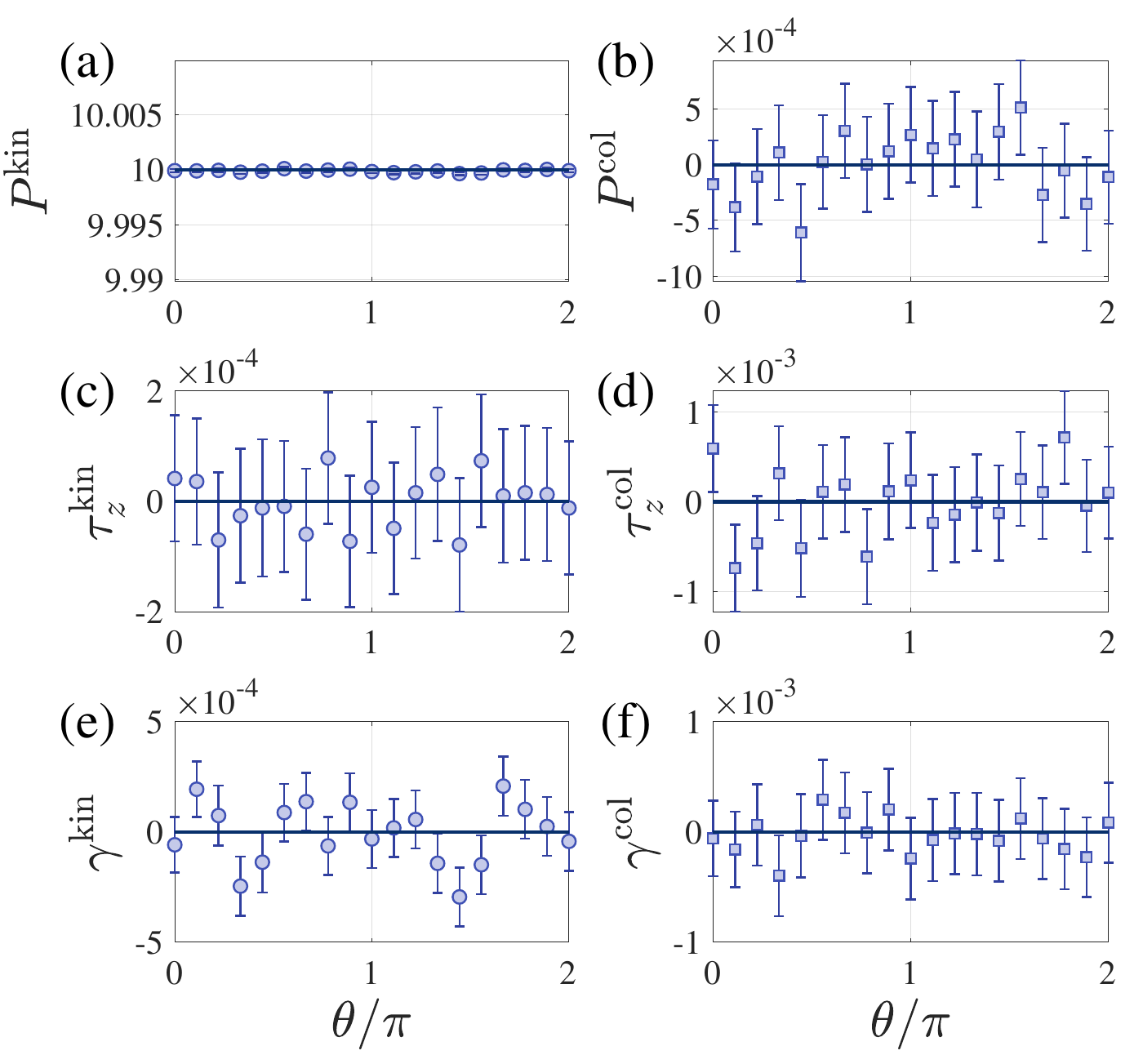}
    \caption{Kinetic and collisional parts of hydrostatic stress measured in simulations with different $\theta$. Apart from the kinetic pressure (a), other parts are approximately zero. The solid line refers to the theoretical value. {  Parameters: $\h=0.1$, $\lambda=10$, $\omega=\pi/3$, $k_BT=1$}. }
    \label{Fig::HydrostaticStress}
\end{figure}

We first test the hydrostatic stress of 3D-CSRD in simulation. The results (see Fig.\ref{Fig::HydrostaticStress}) show that there only exists a kinetic pressure part of the hydrostatic stress. Furthermore, we find the pressure of the 3D-CSRD follows the equation of state for the ideal gas, namely $P=\rho k_BT/m$. This is consistent with our theoretical prediction above (see Eqs.\eqref{HStress_s} and \eqref{Pressure_s}). This consequence allows us to regard the total stresses $\sigma^{2\text{--}9}$ measured in simulations as the viscous stresses ${\sigma^{\rm{v}}}^{2\text{--}9}$, i.e. ${\sigma^{\rm{v}}}^{2\text{--}9}=\sigma^{2\text{--}9}$. For the first viscous stress ${\s^{\rm{v}}}^1=\sqrt{\frac{2}{3}}\left( {\s^{\rm{v}}}_{xx}+{\s^{\rm{v}}}_{yy}+{\s^{\rm{v}}}_{zz} \right)$, we calculate it in simulations by subtracting the pressure from the total stress:
\begin{equation}
    {\s^{\rm{v}}}^1=\sqrt{\frac{2}{3}}\left( {\s}_{xx}+{\s}_{yy}+{\s}_{zz}+3\rho k_BT/m \right).
\end{equation}

\subsection{Viscosities obtained from simulations}
\begin{figure*}[htbp]
    \centering
    \includegraphics[keepaspectratio, width=1.6\columnwidth]{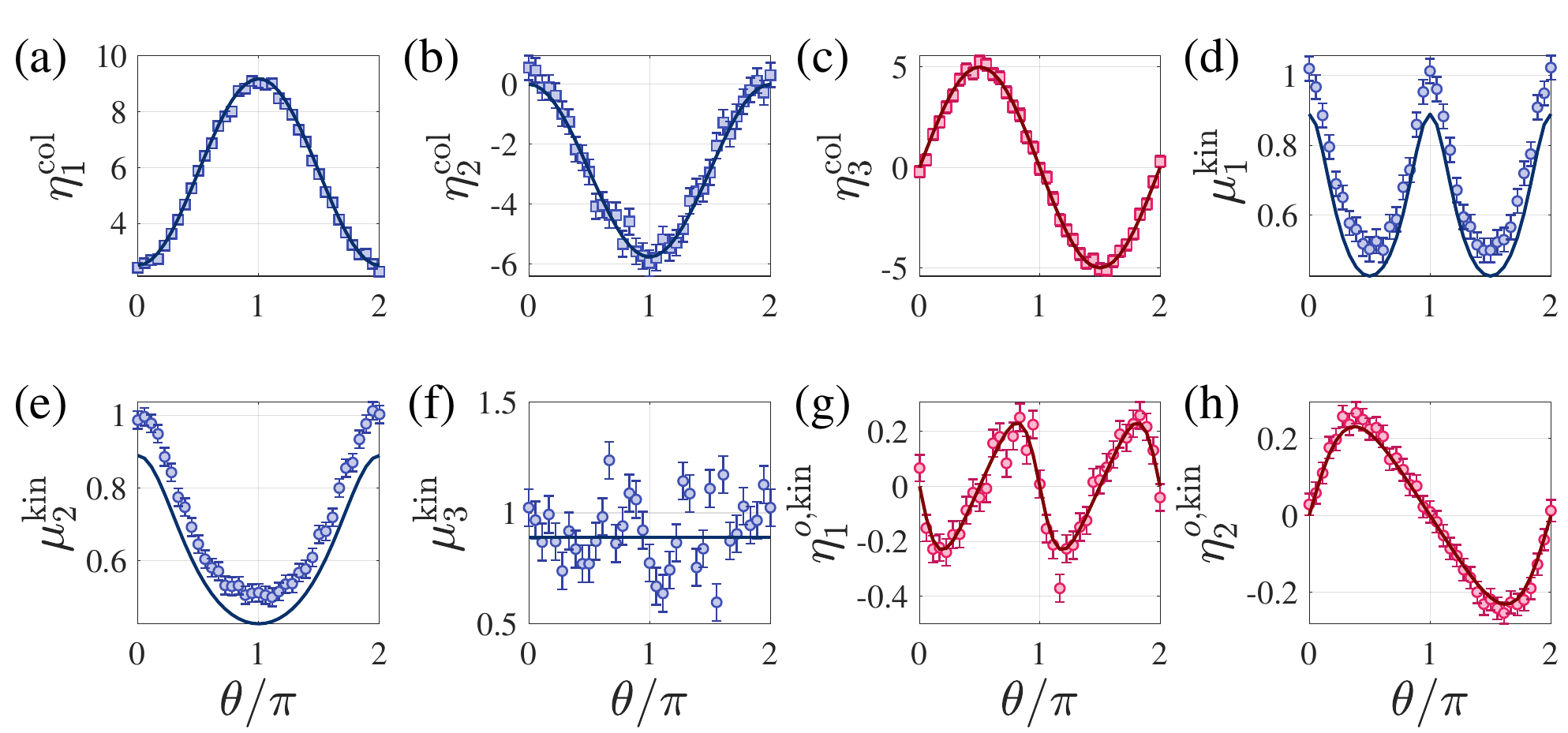}
    \caption{Kinetic and collisional parts of viscosities measured in simulations with different $\theta$. Here, figures (a--c) are the collisional viscosities and figures (d--h) are the kinetic viscosities. The symbols represent the simulation results and the solid lines correspond to the theoretical predictions. {  Parameters: $\h=0.1$, $\lambda=10$, $\omega=\pi/3$, $k_BT=1$.}}
    \label{Fig::ColKinVis}
\end{figure*}

We obtain all of the viscosities of 3D-CSRD fluids via the methods above. The simulation results quantitatively agree with our theoretical predictions. We show three independent components of collisional viscosities (i.e., $\eta_1^{\rm{col}}$, $\eta_2^{\rm{col}}$, and $\eta_3^{\rm{col}}$ in Eqs.~\eqref{ColVis_s}) and five non-zero kinetic viscosities (i.e., $\mu_1^{\rm{kin}}$, $\mu_2^{\rm{kin}}$, $\mu_3^{\rm{kin}}$, $\eta_1^{o,\rm{kin}}$, and $\eta_2^{o,\rm{kin}}$ in Eqs.~\eqref{KinVis_s}) in Fig.~\ref{Fig::ColKinVis}, including both simulation and theoretical results. A slight difference between the simulation data and theoretical values in kinetic viscosities (Figs.~\ref{Fig::ColKinVis}.(d--f)) may arise from the molecular chaos hypothesis employed in the derivation. In Fig.~\ref{Fig::ColKinVis}, the viscosities $\eta_1^{\rm{col}}$, $\eta_2^{\rm{col}}$, and $\eta_3^{\rm{col}}$ are obtained from the data of $\eta_{R,1}^{\rm{col}}$, $\eta_{R,2}^{\rm{col}}$, and $\eta_1^{o,\rm{col}}$ via the following relations according to Eqs.~\eqref{ColVis_s}:
\begin{equation}
    \begin{aligned}
        &\eta_1^{\rm{col}}=\frac{2}{3}\left( 2\eta_{R,1}^{\rm{col}}+\eta_{R,2}^{\rm{col}} \right),\\
        &\eta_2^{\rm{col}}=\frac{4}{\sqrt{3}}\left( \eta_{R,1}^{\rm{col}}-\eta_{R,2}^{\rm{col}} \right),\\
        &\eta_3^{\rm{col}}=2\eta_1^{o,\rm{col}}.
    \end{aligned}
\end{equation}
We provide a complete measurement results of all viscosities in the appendix \ref{Appendix}.

\section{Case study: odd planar Poiseuille flow}\label{Sec::CaseStudy}
In order to validate our simulation model and the hydrodynamic equations derived above, we study the Poiseuille flows of odd fluids by means of both simulation and theory in this section. 

We confine the CSRD fluid between a pair of planes separated by a distance $L$, with the no-slip boundary condition, and drive the fluid via a gravity $\bm{g}$ parallel to the planes. The no-slip boundary condition is realized by the bounce-back rule on the boundary walls. The gravity is performed on the fluid particles in the streaming step as follows
\begin{equation}
    \begin{aligned}
     r_{\a,i}(t+\h)&=r_{\a,i}(t)+v_{\a,i}(t)\h+\frac{1}{2}g_\a\h^2,\\
     v_{\a,i}(t+\h)&=v_{\a,i}(t)+g_\a\h. 
    \end{aligned}
\end{equation}
Here, we again use the Maxwell-Boltzmann scaling thermostat in the simulation to keep the temperature fixed.

\begin{figure*}[htbp]
    \centering
    \includegraphics[keepaspectratio, width=1.6\columnwidth]{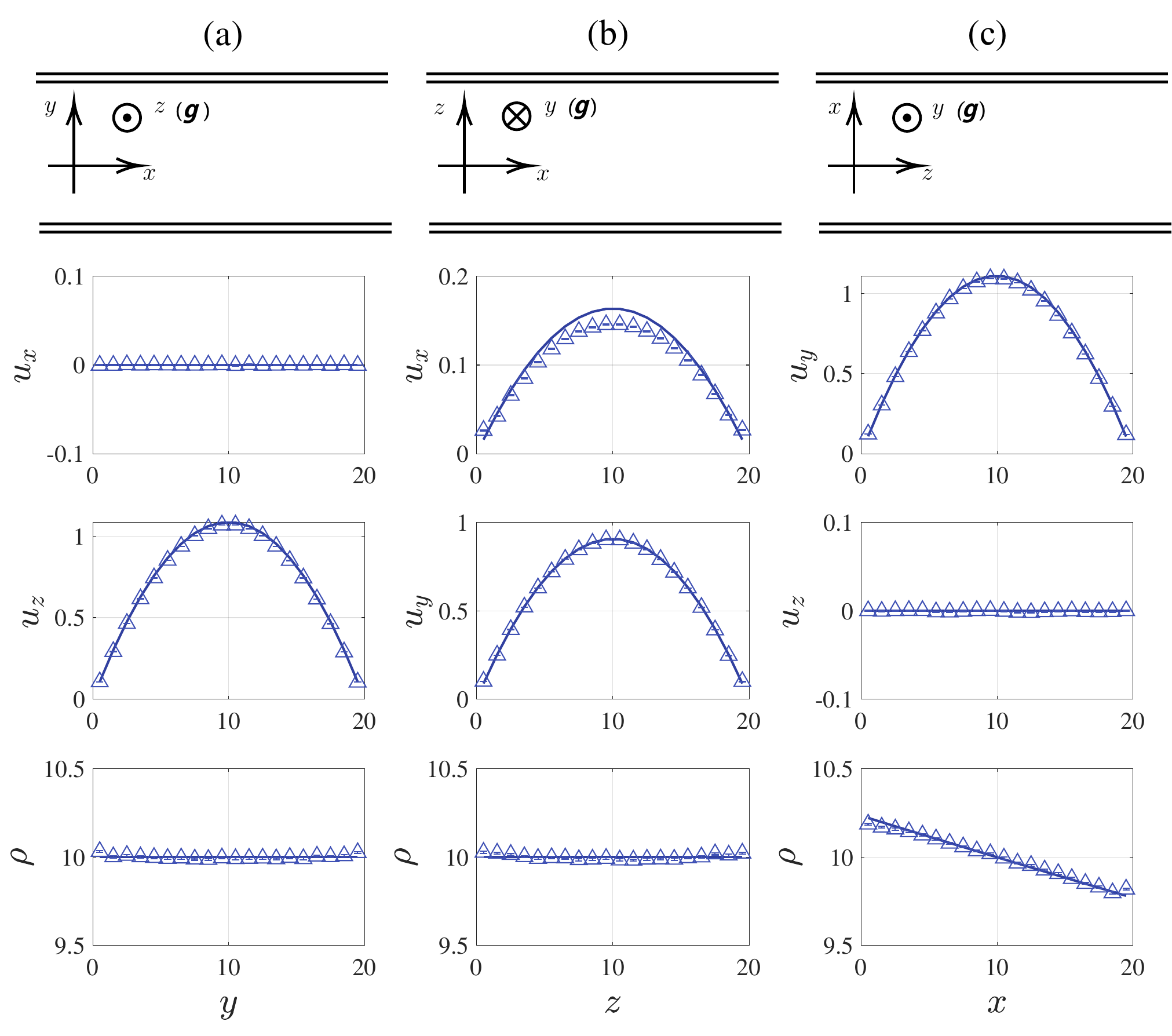}
    \caption{Planar Poiseuille flows in 3D odd fluids. (a) $\bm{g}=g\hat{e}_z$ and the fluid is confined in the $y$ direction. (b) $\bm{g}=g\hat{e}_y$ and the fluid is confined in the $z$ direction. (c) $\bm{g}=g\hat{e}_y$ and the fluid is confined in the $x$ direction. The parameters of the CSRD are set to $\h=0.8$, $\lambda=10$, $k_BT=1$, $\omega=\pi/3$, and $\theta=5\pi/9$. The simulations are performed in a cubic box of dimension $L=20$, and the periodic boundary conditions are applied to the unconfined directions. The symbols $\triangle$ represent the simulation data and the solid lines are the corresponding theoretical predictions. }
    \label{Fig::Poi}
\end{figure*}

Three distinct scenarios of planar Poiseuille flow in the 3D odd fluids are illustrated by the sketches in Fig.~\ref{Fig::Poi}, classified according to the direction of $\bm{g}$ and the position of the boundary walls. In system (a) (Fig.~\ref{Fig::Poi}(a)), the gravity is along the $z$-axis. In system (b) (Fig.~\ref{Fig::Poi}(b)), the gravity is not along the $z$-axis, while the fluid is confined in the $z$ direction. In system (c) (Fig.~\ref{Fig::Poi}(c)), the gravity is not along the $z$-axis and the fluid is also not confined in the $z$ direction. The simulation results are given in Fig.~\ref{Fig::Poi}. For comparison, we analytically calculate these flows by solving the Navier-Stokes equation (Eq.~\eqref{NSEq_s}).

The control equations of flow in system (a) are derived as:
\begin{equation}
    \begin{aligned}
        \eta_{a}\partial_y^2u_z&=-\rho g,\\
        \partial_y^2u_x&=0,\\
        \pd{y}P&=0,
    \end{aligned}
\end{equation}
where $\eta_{a}=\eta_{R,1}-2\eta_{Q,1}^e+\mu_2$ and $P=\rho k_BT/m$. Combining the no-slip boundary condition and $u_y=0$, we obtain the solution:
\begin{equation}
    \begin{aligned}
        u_x&=0,\\
        u_z&=\frac{\rho g}{2\eta_a}y\left( L-y \right).
    \end{aligned}
\end{equation}
This flow is totally consistent with the normal fluids and not affected by the odd viscosities. The profile of $u_z$ and the corresponding simulation result are plotted in Fig.~\ref{Fig::Poi}.(a). However, in systems (b) and (c), situations are different.

For the flow in system (b), we have the following control equations:
\begin{equation}
    \begin{aligned}
        \eta_b\partial_z^2u_x+\eta_{o,b}\partial_z^2u_y&=0,\\
        \eta_b\partial_z^2u_y-\eta_{o,b}\partial_z^2u_x&=-\rho g,\\
        \pd{z}P&=0,
    \end{aligned}
\end{equation}
where $\eta_b=\eta_{R,1}+2\eta_{Q,1}^e+\mu_2$, $\eta_{o,b}=\eta_{R}^o-2\eta_{Q,2}^o-\eta_2^o$, and $P=\rho k_BT/m$. Also applying no-slip boundary condition and $u_z=0$, we have
\begin{equation}
    \begin{aligned}
        u_x&=-\frac{\lambda_b\rho g}{2\left( \eta_b+\lambda_b\eta_{o,b} \right)}z\left( L-z \right),\\
        u_y&=\frac{\rho g}{2\left( \eta_b+\lambda_b\eta_{o,b} \right)}z\left( L-z \right),        
    \end{aligned}
\end{equation}
with $\lambda_b=\eta_{o,b}/\eta_b$. We note, in this condition, the viscosities induce a flow perpendicular to the gravity $\bm{g}$ (see Fig.~\ref{Fig::Poi}.(b)).

The control equations of the flow in system (c) is
\begin{equation}
    \begin{aligned}
        \eta_{o,c}\partial_x^2u_y&=\pd{x}P,\\
        \eta_c\partial_x^2u_y&=-\rho g,\\
        \partial_x^2u_z&=0,
    \end{aligned}
\end{equation}
where $\eta_c=\eta_{R,2}+\mu_1,\,\eta_{o,c}=\eta_1^o-2\eta_{Q,2}^o$, and $P=\rho k_BT/m$. Here the density field $\rho=\rho\left( x \right)$ obeys the normalizing condition $\int_0^L \rho dx=\rho_0L$ with $\rho_0$ the average density. The solution of this flow is:
\begin{equation}
    \begin{aligned}
        \rho&=\frac{\gamma_c\rho_0L}{1-e^{-\gamma_cL}}e^{-\gamma_cx},\\
        u_y&=\frac{\rho_0gL}{\eta_c\gamma_c}\left( \frac{1-e^{-\gamma_cx}}{1-e^{-\gamma_cL}} -\frac{x}{L}\right),\\
        u_z&=0,
    \end{aligned}
\end{equation}
where $\gamma_c=\frac{m\eta_{o,c}g}{\eta_c k_BT}$. The velocity and density profiles are shown in Fig.~\ref{Fig::Poi}.(c). In this case, the odd viscosities give rise to a Hall-like (transverse) mass transport which is also consistent with the phenomenon in 2D odd Poiseuille flow studied in our previous work~\cite{CSRD}.

The excellent agreement between theory and simulations confirms that the derived hydrodynamic equations accurately describe the 3D-CSRD fluids. These results also demonstrate that the 3D mesoscale fluid exhibits rich and abnormal transport phenomena caused by its odd transport coefficients.

\section{Conclusion}\label{Sec::Conclusion}
In this work, we develop a mesoscale simulation model for 3D odd fluids (3D-CSRD). We have demonstrated that this model correctly captures the viscosities and hydrodynamics of odd fluids with $C_\infty$ cylindrical symmetry through theoretical derivation and simulation. Thus, the 3D-CSRD provides a highly efficient platform for studies on intricate hydrodynamics of 3D odd fluids.

With the help of the 3D-CSRD approach, simulations of odd complex fluids can be directly performed. Since the CSRD is an extension of the well-established SRD, the simulation techniques for complex fluids used in SRD are also applicable to the CSRD. For example, to simulate odd colloidal suspensions with CSRD, we can directly employ the hybrid MD-SRD method~\cite{MPC_MD}—a method for simulating mesoscopic suspensions—in which interactions between colloidal particles and fluid particles are properly taken into account, and the colloidal equations of motion are solved in the streaming step via the velocity Verlet algorithm. So, the proposed CSRD approach paves the way for investigating odd complex fluids.  

\qquad\newline
\qquad\newline
\section*{Acknowledgment}
This work was supported by the National Natural Science Foundation of China (No. T2325027, No. 12274448).

\qquad\newline
\qquad\newline
\appendix

\section{Theoretical derivation of Navier-Stokes equation and viscosities}\label{Appendix::KineticDerivation}
We derive the Navier-Stokes equation of the 3D-CSRD by using a kinetic approach proposed by Pooley and Yeomans in their derivation \cite{Pooley} of the hydrodynamic equations for the original SRD. 

We denote the single-particle distribution function of CSRD by $f\left( \bm{r},\bm{v} \right)$ which is normalized by $\int\df\bm{r}\df\bm{v}f=m\mathcal{N}$. This gives the mass density $\rho\left( \bm{r} \right)=\int\df\bm{v}f$. Proceeding, the distribution of a quantity $X=X\left( \bm{r},\bm{v} \right)$ on space is defined by $\bk{X\left( \bm{r},\bm{v} \right)}\triangleq\frac{1}{\rho}\int\df\bm{v}Xf$. In particular, the flow field $\bm{u}\left( \bm{r} \right)$ is defined as $\bm{u}\left( \bm{r} \right)=\bk{\bm{v}}$. Define the velocity moments as $M_{\a\b\cdots}\triangleq\bk{\left( v_\a-u_\a \right)\left( v_\b-u_\b \right)\cdots}$. The second moment is related to the temperature by $\T\triangleq k_BT/m=\frac{1}{3}M_{\a\a}$. We focus on the hydrodynamic behavior of the CSRD fluid in the near-equilibrium state. So, we assume that the fluid is in a local thermodynamic equilibrium and the conserved quantities vary slowly in time and space. These allow us to take the $n$th-order gradients of the conserved quantities as small quantities of magnitude $\mathcal{O}\left( \delta^n \right)$, and express $f\left( \bm{r},\bm{v} \right)$ via local thermodynamic quantities as
\begin{equation}\label{localEqdistribution}
    f(\bm{r},\bm{v})=\frac{\rho(\bm{r})}{\Tp{3/2}(\bm{r})}g\left( \frac{\bm{v}-\bm{u}(\bm{r})}{\sqrt{\T(\bm{r})}} \right)
\end{equation}
with $g(x)$ being a function of the dimensionless quantity $x$.

The conservation laws of mass, momentum, and energy lead to the hydrodynamics. We denote the conserved quantity by $Q=Q\left( \bm{v} \right)$ and the corresponding density and flux by $\rho_Q=\bk{Q}$ and $\bm{J}^{(Q)}$. Then the general form of conservation equation is
\begin{equation}\label{Q}
    \pd{t}\rho_Q+\pd{\a}J_\a^{(Q)}=0.
\end{equation}
However, the situation is of a subtle difference for the CSRD because of its discrete-time dynamics. The flux of $Q$ in the CSRD (denoted by $\bm{j}^{(Q)}$) is the ``discrete flux'' and should be treated as an average of the flux $\bm{J}^{(Q)}$ during $\h$:
\begin{equation}
    j_\a^{(Q)}(t)=\frac{1}{\h}\int_{t}^{t+\h}dt^\prime J_\a^{(Q)}(t^\prime)=J_\a^{(Q)}\left( t+\tau \right),
\end{equation}
where $\tau\in\left[ 0,\h \right]$. This relation gives $J_\a^{(Q)}\left( t \right)=j_\a^{(Q)}\left( t-\tau \right)$. We expand this at $t$ to $\mathcal{O}\left( \delta \right)$, take $\tau=\h/2$ for approximation and then obtain 
\begin{equation}\label{Flux_DtoC}
    J_\a^{(Q)}\left( t \right)=j_\a^{(Q)}\left( t \right)-\frac{\h}{2}\pd{t}j_\a^{(Q)}\left( t \right)+\mathcal{O}\left( \delta^2 \right).
\end{equation}
Both of the streaming step and the collision step contribute to the discrete flux $j_\a^{(Q)}$. We name these two parts by the \emph{kinetic part} and the \emph{collisional part}, respectively, and use the superscript $\rm{kin}$ and $\rm{col}$ to label them: $j_\a^{(Q)}=j_\a^{(Q),\rm{kin}}+j_\a^{(Q),\rm{col}}$. With the help of the local equilibrium distribution Eq.~\eqref{localEqdistribution} and the CSRD dynamics in Eqs.~\eqref{strEq} and \eqref{colEq}, the specific expression of flux $j_\a^{(Q)}$ can be derived and then the hydrodynamic equation can be obtained.

\subsection{Derivation of the collisional flux}
The flux in the collision step can be easily derived using the conservation of $Q$ in each collision cell. We consider a cell divided by a plane $y=c$. The average change of $Q$ during a collision in the upper-half of the cell corresponds to the discrete collisional flux $j_y^{(Q),\rm{col}}$. If we choose the center of this cell as the origin, the discrete flux can be written as:
\begin{equation}
    \begin{aligned}
        j_y^{(Q),\rm{col}}&=\ave{\frac{1}{l^2\h}\int_{-l/2}^{l/2}dz\int_{-l/2}^{l/2}dx\cdot\right.\\
        &\left.\int_{c}^{l/2}dy\rho\bk{Q\left( \bm{v}^c \right)/m-Q\left( \bm{v} \right)/m}},
    \end{aligned}
\end{equation}
where $\bm{v}^c$ is the velocity of a particle after collision. Note that here the position of the plane $c$ follows a uniform distribution in $\left[ -l/2,l/2 \right]$ (i.e., $c\sim U(-l/2,l/2)$) because of the random shift of the lattice. Generally, we have
\begin{equation}\label{Colj}
    \begin{aligned}
        j_\a^{(Q),\rm{col}}&=\ave{\frac{1}{l^2\h}\prod_{\b\neq\a}\left( \int_{-l/2}^{l/2}dx_\b \right)\cdot\right.\\
        &\left.\int_{c}^{l/2}dx_\a\rho\bk{Q\left( \bm{v}^c \right)/m-Q\left( \bm{v} \right)/m}}.
    \end{aligned}
\end{equation}
The mass transport is absent from the collision step so that the collisional mass flux is zero: $J_\a^{(m),\rm{col}}=j_\a^{(m),\rm{col}}=0$.

It is useful to introduce a ``single particle collision formula'' from the collision rule Eq.~\eqref{colEq} to derive the collisional momentum flux. We select one particle in a cell with $N\geqslant 1$ particles and denote its position and velocity before the collision by $\bm{r}$ and $\bm{v}$ respectively. Then we define the mean velocity of other particles by $\hat{\bm{v}}$. Thus, the center-of-mass velocity of this cell is
\begin{equation}\label{Vcm}
    v_{cm,\a}=\frac{1}{N}v_\a+\frac{N-1}{N}\hat{v}_\a.
\end{equation}
Using Eqs.~\eqref{colEq} and \eqref{Vcm}, we can express the velocity after collision of this particle by the following ``single particle collision formula'':
\begin{equation}\label{SingleColEq}
    \begin{aligned}
        v_\a^c&=v_\a+\frac{N-1}{N}\left( R_{\a\b}-\kr{\a\b} \right)\left( v_\b-\hat{v}_\b \right)\\
        &\triangleq v_\a-L_{\a\b}\left( v_\b-\hat{v}_\b \right).
    \end{aligned}
\end{equation}
The particle number in a cell is assumed to follow a Poisson distribution with the expectation value $\n$. Therefore, the probability for $N=q$ particles in the cell containing our selected particle is $P(N=q)=e^{-\n}\n^{q-1}/(q-1)!,q\geqslant1$. Providing more information about the rotation matrix here is necessary. An arbitrary rotation matrix $\tilde{R}_{\a\b}=\tilde{R}_{\a\b}\left( \tilde{\bm{n}},\phi \right)$ can be represented by
\begin{equation}
    \tilde{R}_{\a\b}\left( \tilde{\bm{n}},\phi \right)=\cos\phi\kr{\a\b}+\left( 1-\cos\phi \right)\tilde{n}_\a \tilde{n}_\b-\sin\phi \tilde{n}_\g \ep{\g\a\b}.
\end{equation}
By applying this we have the explicit forms of $R_{\a\b}^1$ and $R_{\a\b}^2$:
\begin{equation}\label{Rot_1}
        R_{\a\b}^1=\cos\a\kr{\a\b}+\left( 1-\cos\a \right)n_\a n_\b-\sin\a n_\g \ep{\g\a\b},
\end{equation}
\begin{equation}\label{Rot_2}
    R_{\a\b}^2=\cos\theta\kr{\a\b}+\left( 1-\cos\theta \right)\kr{z\a}\kr{z\b}-\sin\theta \ep{z\a\b},
\end{equation}
where the random rotation axis $\bm{n}$ is uniformly distributed on the unit sphere. Next, we can calculate $R_{\a\b}=R_{\a\g}^2R_{\g\b}^1$ and obtain:
\begin{equation}
    \begin{aligned}
        R_{\a\b}&=\cos\a\cos\theta\kr{\a\b}+\left( 1-\cos\a \right)\cos\theta n_\a n_\b\\
                &-\sin\a\cos\theta n_\g \ep{\g\a\b}+\cos\a\left( 1-\cos\theta \right)\kr{z\a}\kr{z\b}\\
              &+\left( 1-\cos\a \right)\left( 1-\cos\theta \right)\kr{z\a}n_z n_\b\\
              &+\sin\a\left( 1-\cos\theta \right)\kr{z\a}n_\g\ep{z\g\b}\\
              &-\cos\a\sin\theta\ep{z\a\b}-\left( 1-\cos\a \right)\sin\theta\ep{z\a\g}n_\g n_\b\\
              &+\sin\a\sin\theta\left( n_\a\kr{z\b}-n_z\kr{\a\b} \right).
    \end{aligned}
\end{equation}
The average of $R_{\a\b}$ is:
\begin{equation}\label{aveR}
    \begin{aligned}
        &\ovl{R}_{\a\b}\triangleq\ave{R_{\a\b}}\\
        &=\frac{1}{3}\left( 1+2\cos\a \right)\left[ \cos\theta\kr{\a\b}+\left( 1-\cos\theta \right)\kr{z\a}\kr{z\b}\right.\\
        &\qquad\qquad\qquad\qquad\left.-\sin\theta\ep{z\a\b} \right].
    \end{aligned}
\end{equation}

Now, we set $Q\left( \bm{v} \right)=p_\a$ in Eq.~\eqref{Colj} to calculate the discrete momentum flux $j_\b^{(p_\a),\rm{col}}$:
\begin{equation}\label{ColTab_D1}
    \begin{aligned}
        j_\b^{(p_\a),\rm{col}}&=\ave{\frac{1}{l^2\h}\prod_{\mu\neq\b}\left( \int_{-l/2}^{l/2}dx_\mu \right)\cdot\right.\\
        &\left.\int_{c}^{l/2}dx_\b\rho\bk{v_\a^c-v_\a}}.
    \end{aligned}
\end{equation}
The average change of the velocity $\ave{\bk{v_\a^c-v_\a}}$ in Eq.~\eqref{ColTab_D1} can be derived by taking average of Eq.~\eqref{SingleColEq}:
\begin{equation}
    \begin{aligned}
        &\ave{\bk{v_\a^c-v_\a}}\\
        &=\ave{\frac{N-1}{N}}\left( \ovl{R}_{\a\b}-\kr{\a\b} \right)\left( u_\b-u_{0,\b} \right)\\
        &=\ave{\frac{N-1}{N}}\left( \ovl{R}_{\a\b}-\kr{\a\b} \right)r_\g\pd{\g}u_\b+\mathcal{O}\left( \delta^2 \right).
    \end{aligned}
\end{equation}
In the first equality, $\hat{v}_\a$ is averaged with respect to the velocity and the position of all other particles and thus the result is the flow velocity at the center of the cell $\bm{u}_0$. After calculating the integral in \eqref{ColTab_D1}, we obtain $j_\b^{(p_\a),\rm{col}}$:
\begin{equation}\label{ColTab_D}
    j_\b^{(p_\a),\rm{col}}=\frac{m}{12l\h}\left( \n-1+e^{-\n} \right)\left( \ovl{R}_{\a\mu}-\kr{\a\mu} \right)\kr{\b\nu}\pd{\nu}u_\mu.
\end{equation}
We note the $j_\b^{(p_\a),\rm{col}}$ is $\mathcal{O}\left( \delta \right)$, so according to Eq.~\eqref{Flux_DtoC} the collisional momentum flux has the same form of Eq.~\eqref{ColTab_D}:
\begin{equation}\label{ColTab_C}
    \begin{aligned}
        &J_\b^{(p_\a),\rm{col}}\triangleq T_{\a\b}^{\rm{col}}\\
        &=\frac{m}{12l\h}\left( \n-1+e^{-\n} \right)\left( \ovl{R}_{\a\mu}-\kr{\a\mu} \right)\kr{\b\nu}\pd{\nu}u_\mu+\mathcal{O}\left( \delta^2 \right),
    \end{aligned}
\end{equation}
where we use $T_{\a\b}$ to represent the momentum flux. Hence, we have the collisional stress $\sigma_{\a\b}^{\rm{col}}=-T_{\a\b}^{\rm{col}}$. The collisional stress only depends on the velocity gradients so that the collisional hydrostatic stress is zero:
\begin{equation}\label{HStress_col}
    \sigma_{\a\b}^{\rm{h,col}}=0.
\end{equation}
Thus, we can write $\sigma_{\a\b}^{\rm{col}}=\sigma_{\a\b}^{\rm{v,col}}$.

\subsubsection{The collisional viscosity}
In Eq.~\eqref{ColTab_C}, we can identify the collisional viscosity as follows:
\begin{equation}\label{eta_col}
    \eta_{\a\b\mu\nu}^{col}=\frac{m}{12l\h}\left( \n-1+e^{-\n} \right)\left( \kr{\a\mu}-\ovl{R}_{\a\mu} \right)\kr{\b\nu}.
\end{equation}
Substituting the average of the rotation matrix (see Eq.~\eqref{aveR}) into this expression and rearranging it, we have
\begin{equation}\label{colvisTensor}
    \eta_{\a\b\mu\nu}^{col}=\eta_1^{\rm{col}}\kr{\a\mu}\kr{\b\nu}+\eta_2^{\rm{col}}\tau_{\a\mu}^7\kr{\b\nu}+\eta_3^{\rm{col}}\ep{z\a\mu}\kr{\b\nu},
\end{equation}
where
\begin{equation}\label{colVis_origin}
    \begin{aligned}
        &\eta_1^{\rm{col}}=\frac{m\left( \n-1+e^{-\n} \right)}{108l\h}\left[ 9 - \left( 1+2\cos\omega \right)\left( 1+2\cos\theta \right) \right],\\
        &\eta_2^{\rm{col}}=-\frac{\sqrt{3}m\left( \n-1+e^{-\n} \right)}{108l\h}\left( 1+2\cos\omega \right)\left( 1-\cos\theta \right),\\
        &\eta_3^{\rm{col}}=\frac{m\left( \n-1+e^{-\n} \right)}{36l\h}\left( 1+2\cos\omega \right)\sin\theta.
    \end{aligned}
\end{equation}
By using the orthogonality relations Eqs.~\eqref{Orth} in Eq.~\eqref{eta_col}, the collisional viscosity tensor represented by basis $\left\{ \psi^I \right\}$ is given by:
\begin{widetext}
    \begin{equation}\label{colVis_Pis}
        \bm{\eta}^{\rm{col}}=\left[
            \begin{smallmatrix}
                \eta_1^{\rm{col}} & 0 & 0 & -\sqrt{\frac{2}{3}}\eta_3^{\rm{col}} & 0 & 0 & \sqrt{\frac{2}{3}}\eta_2^{\rm{col}} & 0 & 0 \\ 
                0 & \eta_1^{\rm{col}}+\frac{1}{2\sqrt{3}}\eta_2^{\rm{col}} & \frac{1}{2}\eta_3^{\rm{col}} & 0 & 0 & 0 & 0 & -\frac{\sqrt{3}}{2}\eta_2^{\rm{col}} & -\frac{1}{2}\eta_3^{\rm{col}} \\ 
                0 & -\frac{1}{2}\eta_3^{\rm{col}} & \eta_1^{\rm{col}}+\frac{1}{2\sqrt{3}}\eta_2^{\rm{col}} & 0 & 0 & 0 & 0 & -\frac{1}{2}\eta_3^{\rm{col}} & \frac{\sqrt{3}}{2}\eta_2^{\rm{col}} \\ 
                \sqrt{\frac{2}{3}}\eta_3^{\rm{col}} & 0 & 0 & \eta_1^{\rm{col}}-\frac{1}{\sqrt{3}}\eta_2^{\rm{col}} & 0 & 0 & -\frac{1}{\sqrt{3}}\eta_3^{\rm{col}} & 0 & 0 \\ 
                0 & 0 & 0 & 0 & \eta_1^{\rm{col}}-\frac{1}{\sqrt{3}}\eta_2^{\rm{col}} & \eta_3^{\rm{col}} & 0 & 0 & 0 \\ 
                0 & 0 & 0 & 0 & -\eta_3^{\rm{col}} & \eta_1^{\rm{col}}-\frac{1}{\sqrt{3}}\eta_2^{\rm{col}} & 0 & 0 & 0 \\ 
                \sqrt{\frac{2}{3}}\eta_2^{\rm{col}} & 0 & 0 & \frac{1}{\sqrt{3}}\eta_3^{\rm{col}} & 0 & 0 & \eta_1^{\rm{col}}+\frac{1}{\sqrt{3}}\eta_2^{\rm{col}} & 0 & 0 \\ 
                0 & -\frac{\sqrt{3}}{2}\eta_2^{\rm{col}} & \frac{1}{2}\eta_3^{\rm{col}} & 0 & 0 & 0 & 0 & \eta_1^{\rm{col}}+\frac{1}{2\sqrt{3}}\eta_2^{\rm{col}} & -\frac{1}{2}\eta_3^{\rm{col}} \\ 
                0 & \frac{1}{2}\eta_3^{\rm{col}} & \frac{\sqrt{3}}{2}\eta_2^{\rm{col}} & 0 & 0 & 0 & 0 & \frac{1}{2}\eta_3^{\rm{col}} & \eta_1^{\rm{col}}+\frac{1}{2\sqrt{3}}\eta_2^{\rm{col}} \\ 
            \end{smallmatrix}
                \right].
    \end{equation}

Finally, comparing this to the general form of viscosity tensor Eq.~\eqref{VisT_Psi}, we obtain the collisional viscosities of 3D-CSRD listed here:
\begin{equation}\label{colVis_PisComp}
    \begin{aligned}
        &\zeta^{\rm{col}}=\frac{1}{3}\eta_1^{\rm{col}},\qquad \eta_{R,1}^{\rm{col}}=\mu_2^{\rm{col}}=\frac{1}{2}\eta_1^{\rm{col}}+\frac{1}{4\sqrt{3}}\eta_2^{\rm{col}},\\
        &\eta_{R,2}^{\rm{col}}=\mu_1^{\rm{col}}=\frac{1}{2}\eta_1^{\rm{col}}-\frac{1}{2\sqrt{3}}\eta_2^{\rm{col}},\qquad \mu_3^{\rm{col}}=\frac{1}{2}\eta_1^{\rm{col}}+\frac{1}{2\sqrt{3}}\eta_2^{\rm{col}},\qquad \eta_{s}^{e,\rm{col}}=-\frac{2\sqrt{2}}{3}\eta_{Q,1}^{e,\rm{col}}=\frac{1}{\sqrt{6}}\eta_2^{\rm{col}},\\
        &\eta_{1}^{o,\rm{col}}=-2\eta_{2}^{o,\rm{col}}=2\eta_{R}^{o,\rm{col}}=-2\eta_{Q,2}^{o,\rm{col}}=\frac{1}{2}\eta_3^{\rm{col}},\qquad \eta_A^{o,\rm{col}}=-\sqrt{2}\eta_{Q,3}^{o,\rm{col}}=\frac{1}{\sqrt{6}}\eta_3^{\rm{col}},\\
        &\eta_A^{e,\rm{col}}=\eta_{Q,2}^{e,\rm{col}}=\eta_{Q,3}^{e,\rm{col}}=\eta_{s}^{o,\rm{col}}=\eta_{Q,1}^{o,\rm{col}}=0.
    \end{aligned}
\end{equation}
It can be verified that the even collisional viscosities are positive definite, indicating that 3D-CSRD exhibits a correct dissipative process in the collision step.
\end{widetext}

\subsection{Derivation of the kinetic flux}
Without loss of generality, we calculate the $y$-component of the discrete kinetic flux at the origin $j_{0y}^{(Q),\rm{kin}}$ (in the following, we use the subscript $0$ to label the value of quantities at the origin). This can be written as the flux across the area centered at the origin $\mathcal{D}=\left\{ \left( x,y,z \right) | \left\lvert x \right\rvert,\left\lvert z \right\rvert\leqslant \frac{a}{2},\,y=0 \right\}$ during $\h$:
\begin{equation}\label{jy_kin}
    j_{0y}^{(Q),\rm{kin}}=\frac{1}{a^2\h}\IntV\int_{A}d\bm{r}Q\left( \bm{v} \right)f/m.
\end{equation}
Here, the domain of the integration $A$ is determined by the trajectory of a particle with velocity $\bm{v}$ that intersects with the area $\mathcal{D}$, which gives
\begin{equation}
    \begin{aligned}
        A=&\left\{ \left( r_x,r_y,r_z \right) | -v_y\h\leqslant r_y\leqslant 0,\left\lvert r_x-\frac{v_x}{v_y}r_y \right\rvert\leqslant \frac{a}{2},\right.\\
        &\left.\left\lvert r_z-\frac{v_z}{v_y}r_y \right\rvert\leqslant \frac{a}{2} \right\}.
    \end{aligned}
\end{equation}
The integration in Eq.~\eqref{jy_kin} can be calculated by performing the following variable substitution of velocity $\bm{v}\rightarrow\bm{v}^\pr$:
\begin{equation}\label{VChange}
    \frac{\bm{v}^\pr-\bm{u}_0}{\sqrt{\T_0}}=\frac{\bm{v}-\bm{u}}{\sqrt{\T}}.
\end{equation}
and expanding $\rho,\,\bm{u},\,\T$ at the origin to $\mathcal{O}\left( \delta \right)$ as follows:
\begin{align}
    \rho&=\rho_0+r_\a \bko{\pd{\a}\rho}+\mathcal{O}(\delta^2),\\
    u_\b&=u_{0\b}+r_\a \bko{\pd{\a}u_\b}+\mathcal{O}(\delta^2),\\
    \T&=\T_0+r_\a \bko{\pd{\a}\T}+\mathcal{O}(\delta^2),
\end{align}
where $u_{0\b}$ is also treated as $\mathcal{O}\left( \delta \right)$, which is achievable by choosing a suitable reference frame. Then Eq.~\eqref{jy_kin} becomes
\begin{equation}\label{Epjy_kin}
    \begin{aligned}
        j_{0y}^{(Q),\rm{kin}}&=\frac{1}{a^2\h}\IntV^\pr f(\bm{0},\bm{v}^\pr)\cdot\\
        &\int_{A^\pr}d\bm{r}\frac{1}{m}Q\left(\bm{v}\right)\left[ 1+\frac{1}{\rho_0}r_\a\bko{\pd{\a}\rho} \right],
    \end{aligned}
\end{equation}
where $\bm{v}=\left( 1+\frac{1}{2\T_0}r_\a\bko{\pd{\a}\T} \right)\vb{v}^\pr+r_\a\bko{\pd{\a}\vb{u}}$ and $A^\pr$ is the transformed integration domain. We retain terms up to $\mathcal{O}\left( \delta \right)$ in the integration of $\bm{r}$ in Eq.~\eqref{Epjy_kin}. Thus, the required integrals are just $I^{(0)}\triangleq\frac{1}{a^2\h}\int_{A^\pr}d\bm{r}$ and $I_\a^{(1)}\triangleq\frac{1}{a^2\h}\int_{A^\pr}d\bm{r} r_\a$, which correspond to the terms of order $\mathcal{O}\left( 1 \right)$ and $\mathcal{O}\left( \delta \right)$ in Eq.~\eqref{Epjy_kin}, respectively. These two integrals are given by \cite{Pooley}:
\begin{equation}\label{Int_0}
    \begin{aligned}
        I^{(0)}=&\,\frac{1}{a^2\h}\int_{A^\pr} d\vb{r}\\
        =&\,v_y^\pr-\frac{1}{2}\h\left\{ \frac{\left( v_y^\pr \right)^2\pdo{y}{\T}}{\T_0}+2v_y^\pr\pdo{y}{u_y}\right.\\
        &\left. +v_y^\pr v_\a^\pr\frac{\pdo{\a}{\T}}{\T_0}+v_y^\pr\pdo{\a}{u_\a}+v_\a^\pr\pdo{\a}{u_y}\right\},
    \end{aligned}
\end{equation}
\begin{equation}\label{Int_1}
    I_\a^{(1)}=\frac{1}{a^2\h}\int_{A^\pr} d\bm{r}r_\a=-\frac{1}{2}\h v_y^\pr v_\a^\pr.
\end{equation}

The discrete kinetic fluxes of mass and momentum can be calculated by setting $Q=m$ and $Q=p_\a$ in Eq.~\eqref{Epjy_kin} respectively:
\begin{equation}\label{jm_D}
    \begin{aligned}
        j_{\a}^{(m),\rm{kin}}=j_\a^{(m)}=&\,\rho u_{\a}-\frac{1}{2}\h\pd{\a}{\left( \rho\T \right)}\\
        &-\frac{1}{2}\h\pd{\b}{\left( \rho u_\b u_\a \right)}+\mathcal{O}(\delta^2),
    \end{aligned}
\end{equation}
\begin{equation}\label{KinTab_D}
    \begin{aligned}
        j_{\b}^{(p_\a),\rm{kin}}=&\,\rho u_\a u_\b+\rho M_{\a\b}-\frac{1}{2}\rho\T\h\left( \pd{\a}u_\b+\pd{\b}u_\a\right.\\
        &\left.+\kr{\a\b}\pd{\gamma}u_\gamma \right)+\mathcal{O}(\delta^2),
    \end{aligned}
\end{equation}
in which we have removed the subscript $0$ to represent the fluxes at arbitrary point. Note that the mass transport only happens in the streaming step, so we have $j_{\a}^{(m),\rm{kin}}=j_\a^{(m)}$. Using Eq.~\eqref{Flux_DtoC}, the mass flux and kinetic momentum flux are derived. The mass flux can be directly calculated:
\begin{equation}\label{jm_C1}
    \begin{aligned}
        J_\a^{(m)}=&\,\rho u_\a-\frac{\h}{2}\left[ \pd{t}\left( \rho u_\a \right) + \pd{\a}\left( \rho\T \right) + \pd{\b}\left( \rho u_\b u_\a \right)\right]\\
        &+\mathcal{O}\left( \delta^2 \right)
    \end{aligned}
\end{equation}
To derive the momentum flux, the second moment $M_{\a\b}$ should be decomposed by $M_{\a\b}=\T\kr{\a\b}+M_{\a\b}^\pr$ with $M_{\a\b}^\pr$ is the traceless part. Then the term of time derivative in Eq.~\eqref{Flux_DtoC} will provide a term $\kr{\a\b}\pd{t}\left( \rho\T \right)$ with $\mathcal{O}\left( \delta \right)$ order and other higher order terms. The total form of the kinetic momentum flux therefore is
\begin{equation}\label{KinTab_C1}
    \begin{aligned}
        &J_{\b}^{(p_\a),\rm{kin}}=T_{\a\b}^{\rm{kin}}\\
        =&\,\rho u_\a u_\b+\rho\T\kr{\a\b}+\rho M_{\a\b}^\pr-\frac{1}{2}\rho\T\h\left( \pd{\a}u_\b\right.\\
        &\left.+\pd{\b}u_\a+\kr{\a\b}\pd{\gamma}u_\gamma \right)-\frac{\h}{2}\kr{\a\b}\pd{t}\left( \rho\T \right)+\mathcal{O}(\delta^2).
    \end{aligned}
\end{equation}

These fluxes can be simplified further. Substituting the kinetic and collisional momentum fluxes (Eqs.~\eqref{KinTab_C1} and \eqref{ColTab_C}) into the momentum conservation equation (Eq.~\eqref{Q} with $Q=mv_\a$) yields
\begin{equation}
    \pd{t}\left( \rho u_\a \right)+ \pd{\b}\left( \rho u_\b u_\a \right) + \pd{\a}\left( \rho\T \right)=\mathcal{O}(\delta^2).
\end{equation}
This simplifies the mass flux Eq.~\eqref{jm_C1}:
\begin{equation}\label{jm_C}
    J_\a^{(m)}=\rho u_\a+\mathcal{O}\left( \delta^2 \right).
\end{equation}
Similarly, the kinetic momentum flux will be simplified by considering the energy conservation equation of the $\mathcal{O}\left( \delta \right)$ order. Taking $Q=E_k$ in Eq.~\eqref{Q} gives the energy conservation equation:
\begin{equation}\label{EnergyEq0}
    \pd{t}\left( \frac{1}{2}\rho\bk{v^2} \right)+\pd{\a}q_\a=0,
\end{equation}
where $q_\a$ is the energy flux $q_\a\triangleq J_\a^{E_k}$ and the first term is
\begin{equation}\label{dedt}
    \begin{aligned}
        \pd{t}\left( \frac{1}{2}\rho\bk{\bm{v}^2} \right)&=\pd{t}\left( \frac{3}{2}\rho\T+\frac{1}{2}\rho u^2 \right)&\\
        &=\pd{t}\left( \frac{3}{2}\rho\T \right)+\mathcal{O}(\delta^2).
    \end{aligned}
\end{equation}
Proceeding, we expand the second term: $\pd{\a}q_\a=\pd{\a}q_\a^{\rm{kin}}+\pd{\a}q_\a^{\rm{col}}$. Herein, the collisional term $\pd{\a}q_\a^{\rm{col}}$ is the order of $\mathcal{O}\left( \delta^2 \right)$ at least, as inferred from the derivations in the 2D case~\cite{CSRD}. The discrete kinetic energy flux calculated from Eq.~\eqref{Epjy_kin} is
\begin{equation}\label{Kinqa_D}
    j_{\a}^{(E_k),\rm{kin}}=\frac{5}{2}\rho\T u_\a+\frac{1}{2}\rho M_{\b\b\a}-\frac{5}{4}\h\pd{\a}\left(\rho\Tp{2}\right)+\mathcal{O}(\delta^2).
\end{equation}
Combining Eq.~\eqref{Flux_DtoC}, the kinetic energy flux is
\begin{equation}\label{Kinqa_C}
    \begin{aligned}
        q_\a^{\rm{kin}}=&\,\frac{5}{2}\rho\T u_\a+\frac{1}{2}\rho M_{\b\b\a}-\frac{5}{4}\h\pd{\a}\left(\rho\Tp{2}\right)\\
        &-\frac{\h}{2}\rho\T\pd{t}u_\a+\mathcal{O}(\delta^2).
    \end{aligned}
\end{equation}
Substituting Eqs.~\eqref{dedt} and \eqref{Kinqa_C} into Eq.~\eqref{EnergyEq0}, the energy conservation equation of the $\mathcal{O}\left( \delta \right)$ order is obtained:
\begin{equation}\label{EnergyEq_O1}
    \pd{t}\left( \rho\T \right)=-\frac{5}{3}\rho\T\pd{\a}u_\a+\mathcal{O}\left( \delta^2 \right).
\end{equation}
In the derivation we treat the third momentum $M_{\b\b\a}$ as a quantity with the order of $\mathcal{O}\left( \delta \right)$ at least, because $M_{\b\b\a}$ is exactly zero in the equilibrium state. Using Eqs.~\eqref{KinTab_C1} and \eqref{EnergyEq_O1}, the final expression of the kinetic momentum flux is
\begin{equation}\label{KinTab_C}
    \begin{aligned}
        T_{\a\b}^{\rm{kin}}=&\,\rho u_\a u_\b+\rho\T\kr{\a\b}+\rho M_{\a\b}^\pr\\
        &-\frac{1}{2}\rho\T\h\left( \pd{\a}u_\b+\pd{\b}u_\a-\frac{2}{3}\kr{\a\b}\pd{\gamma}u_\gamma \right)+\mathcal{O}(\delta^2).
    \end{aligned}
\end{equation}
The kinetic stress tensor can be easily read out from the kinetic momentum flux Eq.~\eqref{KinTab_C}:
\begin{equation}
    \begin{aligned}
        \sigma_{\a\b}^{\rm{kin}}=&\,-\rho\T\kr{\a\b}-\rho M_{\a\b}^\pr\\
        &+\frac{1}{2}\rho\T\h\left( \pd{\a}u_\b+\pd{\b}u_\a-\frac{2}{3}\kr{\a\b}\pd{\gamma}u_\gamma \right).
    \end{aligned}
\end{equation}
Referring to Eq.~\eqref{HStress}, the hydrostatic part of kinetic stress only has a pressure term
\begin{equation}\label{HStress_kin}
    \sigma_{\a\b}^{\rm{h,kin}}=-P\kr{\a\b}
\end{equation}
with $P=\rho\T$ since the CSRD follows an ideal gas equation of state in the equilibrium state. The viscous part is
\begin{equation}\label{KinVisStress}
    \sigma_{\a\b}^{\rm{v,kin}}=-\rho M_{\a\b}^\pr+\frac{\rho\T\h}{2}\left( \pd{\a}u_\b+\pd{\b}u_\a-\frac{2}{3}\kr{\a\b}\pd{\gamma}u_\gamma \right).
\end{equation}
The kinetic viscosities are hidden in the second-order moment $M_{\a\b}^\pr$ which are derived below.
\subsubsection{The velocity moment}
During the streaming and collision steps, the single particle distribution $f\left( \bm{r},\bm{v} \right)$ is changed. The velocity moments may also be transformed by these two steps. In general, we can express the transformation relation of the velocity moment during one CSRD step $\left[ t,t+\h \right]$ as an iteration equation $M_{\a\b\cdots}^{t+\h}=\hat{\mathcal{F}}[M_{\a\b\cdots}^t]$. In the steady state, $f\left( \bm{r},\bm{v} \right)$ converges to an invariant distribution, allowing us to calculate the stationary value of the velocity moment by solving the fixed point of the iteration equation:
\begin{equation}\label{FixedP}
    M_{\a\b\cdots}=\hat{\mathcal{F}}[M_{\a\b\cdots}].
\end{equation}
Equation~\eqref{FixedP} can be derived by separately analyzing the streaming and collision operations.

\emph{Transformation in the streaming step.}---Herein, we denote the quantities before and after the streaming step by the superscripts $t$ and $s$ respectively. After being altered by streaming, the single particle distribution becomes 
\begin{equation}
    f^s\left( \bm{r},\bm{v} \right)=f^t\left( \bm{r}-\bm{v}\h,\bm{v} \right).
\end{equation}
Then, the second-order velocity moment after the streaming step is given by
\begin{equation}
    M_{0,\a\b}^s=\frac{1}{\rho^s}\IntV f^{t}(-\bm{v}\h,\bm{v})\left( v_\a-u_\a^s \right)\left( v_\b-u_\b^s \right),
\end{equation}
where we set $\bm{r}=\bm{0}$ for simplicity, with $\rho_0^{s}$ and $u_{0,\a}^s$ being the density and flow velocity after streaming respectively. This integral can be calculated by the same procedure used in the derivation of Eq.~\eqref{jy_kin}. Using the substitution Eq.~\eqref{VChange}, we derive the following transformation relations:
\begin{align}
    &\rho^s=\rho^t-\rho^t\h\pd{\a}u_\a+\mathcal{O}(\delta^2),\label{rhoChange}\\
    &u_\a^s=u_\a^t-\h\frac{1}{\rho^t}\pd{\a}\left( \rho^t\Tp{t} \right)+\mathcal{O}(\delta^2),\label{uaStr}\\
    &M_{\a\b}^s=M_{\a\b}^t-\T\h\left( \pd{\a}u_\b+\pd{\b}u_\a \right)+\mathcal{O}(\delta^2).\label{MabStr}
\end{align}
The traceless part of the second-order moment is therefore transformed by:
\begin{equation}\label{MtToMs}
    M_{\a\b}^{\pr s}=M_{\a\b}^{\pr t}-\T\h\left( \pd{\a}u_{\b}+\pd{\b}u_{\a}-\frac{2}{3}\kr{\a\b}\pd{\g}u_\g \right).
\end{equation}
We can express Eq.~\eqref{MtToMs} in the subspace spanned by bases $\psi^{5\text{--}9}$ via Eqs.~\eqref{BisTran} as follows because $M_{\a\b}^{\pr}$ is traceless and symmetric:
\begin{equation}\label{MtToMs_Psi}
    \begin{bmatrix}
        M^{\pr s,5}\\
        M^{\pr s,6}\\
        M^{\pr s,7}\\
        M^{\pr s,8}\\
        M^{\pr s,9}\\
    \end{bmatrix}
    =
    \begin{bmatrix}
        M^{\pr t,5}\\
        M^{\pr t,6}\\
        M^{\pr t,7}\\
        M^{\pr t,8}\\
        M^{\pr t,9}\\
    \end{bmatrix}
    -2\T\h
    \begin{bmatrix}
        \dot{e}^{5}\\
        \dot{e}^{6}\\
        \dot{e}^{7}\\
        \dot{e}^{8}\\
        \dot{e}^{9}\\
    \end{bmatrix},
\end{equation}
where $\dot{e}^I$ follows the definition in Eqs.~\eqref{Tran_ConstitutiveRelation}.

\emph{Transformation in the collision step.}---To facilitate the calculations in this section, we provide an alternative expression of the collision rule Eq.~\eqref{colEq}. Consider a cell at position $\bm{\xi}$ containing $N$ particles, with velocities denoted as $\bm{v}^{(i)},\,i=1,\dots,N$. Then we arrange these velocities by introducing the vector $\bm{v}_{\bm{\xi}}=\left( \bm{v}^{(1)},\dots,\bm{v}^{(N)} \right)^\top$. Using Eq.~\eqref{colEq} and the vector $\bm{v}_{\bm{\xi}}$, the collision operation in cell $\bm{\xi}$ can be expressed as:
\begin{subequations}\label{cellCol}
    \begin{align}
        &\bm{v}_{\bm{\xi}}\left( t+\h \right)=\bm{\mathcal{C}}_{\bm{\xi}}\cdot\bm{v}_{\bm{\xi}}\left( t \right),\\
        &\bm{\mathcal{C}}_{\bm{\xi}} =
        \left[ \begin{smallmatrix}
            \bm{R}_{\bm{\xi}}+\frac{1}{N}(\bm{I}-\bm{R}_{\bm{\xi}}) & \frac{1}{N}(\bm{I}-\bm{R}_{\bm{\xi}}) & \dots  & \frac{1}{N}(\bm{I}-\bm{R}_{\bm{\xi}})  \\
            \frac{1}{N}(\bm{I}-\bm{R}_{\bm{\xi}}) & \bm{R}_{\bm{\xi}}+\frac{1}{N}(\bm{I}-\bm{R}_{\bm{\xi}}) & \dots  & \frac{1}{N}(\bm{I}-\bm{R}_{\bm{\xi}}) \\
            \vdots  & \vdots   & \ddots & \vdots \\
            \frac{1}{N}(\bm{I}-\bm{R}_{\bm{\xi}}) & \frac{1}{N}(\bm{I}-\bm{R}_{\bm{\xi}})  & \dots  & \bm{R}_{\bm{\xi}}+\frac{1}{N}(\bm{I}-\bm{R}_{\bm{\xi}})
        \end{smallmatrix} \right],
    \end{align}
\end{subequations}
where $\bm{I}$ is the $3\times3$ unit matrix and $\bm{R}_{\bm{\xi}}$ represents the rotation in cell $\bm{\xi}$. The matrix $\bm{\mathcal{C}}_{\bm{\xi}}$ is orthogonal so that $\left\lvert \det\left( \bm{\mathcal{C}}_{\bm{\xi}} \right)\right\rvert = 1$. This indicates the CSRD conserves the phase volume element.

We denote the distribution of a particle (with position $\bm{r}$ and velocity $\bm{v}$) after a collision in the cell $\bm{\xi}=\bm{\xi}\left( \bm{r} \right)$ by $f^{sc}\left( \bm{r},\bm{v} \right)$. The distribution $f^{sc}\left( \bm{r},\bm{v} \right)$ can be expressed by $f^{s}\left( \bm{r},\bm{v} \right)$ as:
\begin{equation}\label{fsc}
    f^{sc}(\bm{r},\bm{v})=\ave{f^{s}(\bm{r},\bm{R}^{-1}_{\bm{\xi}}\cdot\left( \bm{v}-\bm{v}_{cm} \right)+\bm{v}_{cm})}.
\end{equation}
The average in Eq.~\eqref{fsc} is performed over other particles in the same cell as follows:
\begin{equation}\label{fsc_1}
    \begin{aligned}
        &f^{sc}(\bm{r},\bm{v})\\
        &=\ave{f^{s}(\bm{r},\bm{R}^{-1}_{\bm{\xi}}\cdot\left( \bm{v}-\bm{v}_{cm} \right)+\bm{v}_{cm})}\\
        &=\ave{\int d\bm{v}^{(2)}\cdots d\bm{v}^{(N)}f^{s}(\bm{r},\bm{R}^{-1}_{\bm{\xi}}\cdot\left( \bm{v}-\bm{v}_{cm} \right)+\bm{v}_{cm})\cdot\right.\\
        &\qquad \left.p\left( \bm{v}^{(2)},\dots,\bm{v}^{(N)} \mid \bm{r}^{(2)}\in A_{\bm{\xi}},\dots,\bm{r}^{(N)}\in A_{\bm{\xi}} \right)}.
    \end{aligned}
\end{equation}
Here, we use labels $(2),\dots,(N)$ to represent other particles in the cell and define the area of the cell by $A_{\bm{\xi}}$. Next, we apply the molecular chaos hypothesis to decompose the conditional probability by
\begin{equation}
    \begin{aligned}
        &p\left( \bm{v}^{(2)},\dots,\bm{v}^{(N)} \mid \bm{r}^{(2)}\in A_{\bm{\xi}},\dots,\bm{r}^{(N)}\in A_{\bm{\xi}} \right)\\
        &=\prod_{i=2}^N p\left( \bm{v}^{(i)} \mid \bm{r}^{(i)}\in A_{\bm{\xi}} \right).
    \end{aligned}
\end{equation}
Using Bayes' formula the conditional probability of particle $(i)$ reads:
\begin{equation}
    \begin{aligned}
        p\left( \bm{v}^{(i)} \mid \bm{r}^{(i)}\in A_{\bm{\xi}} \right)&=\frac{p\left( \bm{v}^{(i)} ; \bm{r}^{(i)}\in A_{\bm{\xi}} \right)}{p\left( \bm{r}^{(i)}\in A_{\bm{\xi}} \right)}\\
        &=\frac{\int_{A_{\bm{\xi}}}d\bm{r}^{(i)}f^{sc}\left( \bm{r}^{(i)},\bm{v}^{(i)} \right)}{\int_{A_{\bm{\xi}}}d\bm{r}^{(i)}\rho^{s}\left( \bm{r}^{(i)} \right)}\\
        &\approx\frac{f^{sc}\left( \bm{r},\bm{v}^{(i)} \right)}{\rho^{s}\left( \bm{r} \right)}.
    \end{aligned}
\end{equation}
In the last equality, considering the slow variation of distribution in space, we use the values of $f^{sc}$ and $\rho^s$ at $\bm{r}$ ($f^{sc}\left( \bm{r},\bm{v}^{(i)} \right)$ and $\rho^{s}\left( \bm{r} \right)$) to estimate their values in the cell $\bm{\xi}$, which significantly simplifies the derivations below. Therefore, Eq.~\eqref{fsc_1} is simplified as:
\begin{equation}\label{fsc_2}
    \begin{aligned}
        &f^{sc}(\bm{r},\bm{v})\\
        &=\mathbb{E}\left[\frac{1}{\left( \rho^s \right)^{N-1}}\int d\bm{v}^{(2)}\cdots d\bm{v}^{(N)}\right.\\
        &\qquad f^{s}(\bm{r},\bm{R}^{-1}_{\bm{\xi}}\cdot\left( \bm{v}-\bm{v}_{cm} \right)+\bm{v}_{cm})\cdot\\
        &\qquad \left.f^{sc}\left( \bm{r},\bm{v}^{(2)} \right)\cdots f^{sc}\left( \bm{r},\bm{v}^{(N)} \right)\right].
    \end{aligned}
\end{equation}
Then, $f^{sc}$ in Eq.~\eqref{fsc_2} can be substituted by $f^{s}$ via Eq.~\eqref{fsc}:
\begin{equation}\label{fsc_3}
    \begin{aligned}
        &f^{sc}(\bm{r},\bm{v})\\
        &=\mathbb{E}\left[\frac{1}{\left( \rho^s \right)^{N-1}}\int d\bm{v}^{(2)}\cdots d\bm{v}^{(N)}f^{s}_{\bm{\xi}}\left( \bm{r},\bm{v}_{\bm{\xi}}^\pr\right)\right],
    \end{aligned}
\end{equation}
where we have defined the following relations:
\begin{equation}
    \begin{aligned}
        &f^{s}_{\bm{\xi}}\left( \bm{r},\bm{v}_{\bm{\xi}}^\pr\right)\\
        &=f^{s}_{\bm{\xi}}\left( \bm{r},\bm{v}^\pr,\bm{v}^{(2)\pr},\dots ,\bm{v}^{(N)\pr} \right)\\
        &=\prod_{i=1}^{N}f^{s}\left( \bm{r},\bm{v}^{(i)\pr}\right),
    \end{aligned}
\end{equation}
\begin{equation}
    \bm{v}^{(1)\pr}=\bm{v}^{\pr},
\end{equation}
and 
\begin{equation}\label{singleVpr}
    \bm{v}^{(i)\pr}=\bm{R}^{-1}_{\bm{\xi}}\cdot\left( \bm{v}^{(i)}-\bm{v}_{cm} \right)+\bm{v}_{cm}.
\end{equation}
The relation Eq.~\eqref{singleVpr} can be re-written by Eqs.~\eqref{cellCol}:
\begin{equation}
    \bm{v}_{\bm{\xi}}^\pr=\bm{\mathcal{C}}_{\bm{\xi}}^{-1}\cdot\bm{v}_{\bm{\xi}},
\end{equation}
where $\bm{v}_{\bm{\xi}}=\left( \bm{v},\bm{v}^{(2)},\dots ,\bm{v}^{(N)} \right)^\top$. Substituting this into Eq.~\eqref{fsc_3} yields:
\begin{equation}\label{fsc_final}
    \begin{aligned}
        &f^{sc}(\bm{r},\bm{v})\\
        &=\ave{\frac{1}{\left( \rho^s \right)^{N-1}}\int d\bm{v}^{(2)}\cdots d\bm{v}^{(N)}f^{s}_{\bm{\xi}}\left( \bm{r},\bm{\mathcal{C}}_{\bm{\xi}}^{-1}\cdot\bm{v}_{\bm{\xi}} \right)}.
    \end{aligned}
\end{equation}
Eq.~\eqref{fsc_final} allows us to derive any quantities after the transformation of the collision step. First, we can check that the density (the zero-order velocity moment) is invariant in collision:
\begin{equation}
    \rho^{sc}\left( \bm{r} \right)=\IntV f^{sc}\left( \bm{r},\bm{v} \right)=\rho^s\left( \bm{v} \right)
\end{equation}
Other velocity moments after collision are formally expressed by
\begin{equation}
    \begin{aligned}\label{M_sc_1}
        M_{\a\b\cdots}^{sc}=&\frac{1}{\rho^{sc}}\IntV f^{sc}\left( \bm{r},\bm{v} \right)\tilde{M}({\bm{v}})\\
        =&\ave{\frac{1}{\left( \rho^s \right)^{N}}\int d\bm{v_{\vb{\xi}}} f_{\bm{\xi}}^{s}(\bm{r},\bm{\mathcal{C}}_{\bm{\xi}}^{-1}\cdot\bm{v}_{\bm{\xi}})\tilde{M}({\bm{v}})},
    \end{aligned}
\end{equation}
where we set $\tilde{M}({\bm{v}})=v_\a$ to express the first-order moment, i.e., the flow velocity $u_\a^{sc}$, and $\tilde{M}({\bm{v}})=\left( v_\a-u_\a^{sc} \right)\cdot$\\$\left( v_\b-u_\b^{sc} \right)\cdots$ to express other moments. In the calculation of integral in Eq.~\eqref{M_sc_1}, we use the substitution $\bm{v}_{\bm{\xi}}\rightarrow\bm{\mathcal{C}}_{\bm{\xi}}\cdot\bm{v}_{\bm{\xi}}$. Then Eq.~\eqref{M_sc_1} is simplified by
\begin{equation}\label{M_sc_2}
    \begin{aligned}
        M_{\a\b\ldots}^{sc}&=\ave{\frac{1}{\left( \rho^s \right)^{N}}\int d\vb{v}_{\vb{\xi}}f_{\vb{\xi}}^s(\vb{r},\vb{v}_{\vb{\xi}})\tilde{M}\left( \bm{v}-\bm{L}\cdot\left( \bm{v}-\hat{\bm{v}} \right) \right)}\\
        &=\ave{\bk{\tilde{M}\left( \bm{v}-\bm{L}\cdot\left( \bm{v}-\hat{\bm{v}} \right) \right)}^s}\\
        &=\ave{\bk{\tilde{M}\left( \bm{v}^c \right)}^s},
    \end{aligned}
\end{equation}
where $\hat{\bm{v}}$ is the mean velocity of $(2)\text{--}(N)$ particles as defined in Eq.~\eqref{Vcm}, and $\bm{v}^c$ and $\bm{L}$ are the quantities defined in Eq.~\eqref{SingleColEq}, i.e., the ``single particle collision formula'':
\begin{equation}\label{Vc}
    \bm{v}^c=\bm{v}-\bm{L}\cdot\left( \bm{v}-\hat{\bm{v}} \right).
\end{equation}
In the following, we define the notation $\bbkk{\cdot}$ by $\bbkk{\cdot}=\ave{\bk{\cdot}^s}$ for convenience so that Eq.~\eqref{M_sc_2} can be simply written by
\begin{equation}\label{M_sc_final}
    M_{\a\b\ldots}^{sc}=\bbkk{\tilde{M}\left( \bm{v}^c \right)}.
\end{equation}
The first-order moment is also invariant under collision operation:
\begin{equation}
    \begin{aligned}
        u_\a^{sc}&=\bbkk{v_\a^c}=\bbkk{v_\a}-\ave{L_{\a\b}}\bbkk{v_\b-\hat{v}_\b}=\bbkk{v_\a}\\
                 &=u_\a^s.
    \end{aligned}
\end{equation}
Therefore, the second-order moment can be derived by
\begin{equation}\label{Mab}
    M_{\a\b}^{sc}=\bbkk{\left( v_\a^c-u_\a^s \right)\left( v_\b^c-u_\b^s \right)}=\bbkk{v_\a^cv_\b^c}+\mathcal{O}\left( \delta^2 \right).
\end{equation}
Inserting Eq.~\eqref{Vc} into this and using the molecular chaos hypothesis ($\bbkk{v_\a\hat{v}_\b}=0$), we obtain the transform of veloctiy moment in the collision step:
\begin{equation}\label{MsToMsc_1}
    M_{\a\b}^{sc}=\mathcal{L}_{\a\b\mu\nu}^\pr M_{\mu\nu}^s,
\end{equation}
where
\begin{equation}\label{L_prime}
    \mathcal{L}_{\a\b\mu\nu}^\pr=\ave{\frac{1}{N}}\kr{\a\mu}\kr{\b\nu}+\ave{\frac{N-1}{N}}\ave{R_{\a\mu}R_{\b\nu}}.
\end{equation}
Next, we go to calculate the average $\ave{R_{\a\mu}R_{\b\nu}}$ which equals to $R_{\a\g}^2R_{\b\tau}^2\ave{R_{\g\mu}^1R_{\tau\nu}^1}$. Expanding $R_{\a\b}^1$ via Eq.~\eqref{Rot_1} and using the averages about the random rotation axis $\ave{n_\a}=0,\,\ave{n_\a n_\b}=\frac{1}{3}\kr{\a\b},\,\ave{n_\a n_\b n_\g}=0,\,\ave{n_\a n_\b n_\g n_\tau}=\frac{1}{15}\left( \kr{\a\b}\kr{\g\tau}+\kr{\a\g}\kr{\b\tau}+\kr{\a\tau}\kr{\b\g} \right)$ results in,
\begin{equation}
    \begin{aligned}
        \ave{R_{\g\mu}^1R_{\tau\nu}^1}&=q^{(1)}\kr{\g\mu}\kr{\tau\nu}+q^{(2)}\kr{\g\tau}\kr{\mu\nu}+q^{(3)}\kr{\mu\tau}\kr{\g\nu},\\
        q^{(1)}&=\frac{1}{15}\left( 4+8\cos\omega+3\cos 2\omega \right),\\
        q^{(2)}&=\frac{2}{15}\left( 2-\cos\omega-\cos 2\omega \right),\\
        q^{(3)}&=-\frac{1}{15}\left( 1+2\cos\omega-3\cos 2\omega \right).
    \end{aligned}
\end{equation}
Thus, $\mathcal{L}_{\a\b\mu\nu}^\pr$ reads
\begin{equation}
    \begin{aligned}
        \mathcal{L}_{\a\b\mu\nu}^\pr=&\,\ave{\frac{1}{N}}\kr{\a\mu}\kr{\b\nu}+\ave{\frac{N-1}{N}}\left( q^{(1)}R^2_{\a\mu}R^2_{\b\nu}\right.\\
        &\left.+q^{(2)}\kr{\a\b}\kr{\mu\nu}+q^{(3)}R^2_{\a\nu}R^2_{\b\mu} \right).
    \end{aligned}
\end{equation}
In Eq.~\eqref{MsToMsc_1}, the symmetry of the moment ($M_{\a\b}^\pr=M_{\b\a}^\pr$) allow the term of $q^{(3)}$ to be absorbed into the term of $q^{(1)}$. Hence, we rewrite Eq.~\eqref{MsToMsc_1} as:
\begin{equation}\label{MsToMsc_final}
    M_{\a\b}^{sc}=\mathcal{L}_{\a\b\mu\nu}M_{\mu\nu}^s,
\end{equation}
where
\begin{equation}\label{L_sym_1}
    \begin{aligned}
        \mathcal{L}_{\a\b\mu\nu}&=\ave{\frac{1}{N}}\kr{\a\mu}\kr{\b\nu}+\ave{\frac{N-1}{N}}\left( p\kr{\a\b}\kr{\mu\nu}\right.\\
                                &\qquad\left.+qR^2_{\a\mu}R^2_{\b\nu} \right),\\
        p&\triangleq q^{(2)}=\frac{2}{15}\left( 2-\cos\omega-\cos 2\omega \right),\\
        q&\triangleq q^{(1)}+q^{(3)}=\frac{1}{5}\left( 1+2\cos\omega+2\cos 2\omega \right).
    \end{aligned}
\end{equation}
Here the coefficients $p$ and $q$ have the following quantitative relation:
\begin{equation}\label{pq}
    3p+q=1.
\end{equation}
We use Eq.~\eqref{Rot_2} to replace the additional rotation matrices $R^2_{\a\b}$ in $\mathcal{L}_{\a\b\mu\nu}$ and then obtain
\begin{equation}\label{L_eab}
    \begin{aligned}
        &\mathcal{L}_{\a\b\mu\nu}\\
        =&\,\left( \ave{\frac{1}{N}}+q\ave{\frac{N-1}{N}}s^{(1)} \right)\kr{\a\mu}\kr{\b\nu}\\
        &+p\ave{\frac{N-1}{N}}\kr{\a\b}\kr{\mu\nu}+q\ave{\frac{N-1}{N}}\left[ {s^{(2)}} \left( \kr{\a\mu}\tau_{\b\nu}^7\right.\right.\\
        &\left.+\tau_{\a\mu}^7\kr{\b\nu} \right)+{s^{(3)}}\left( \kr{\a\mu}\ep{z\b\nu}+\ep{z\a\mu}\kr{\b\nu} \right)+{s^{(4)}}\left( \tau^7_{\a\mu}\ep{z\b\nu}\right.\\
        &\left.\left.+\ep{z\a\mu}\tau^7_{\b\nu} \right)+{s^{(5)}}\tau^7_{\a\mu}\tau^7_{\b\nu}+{s^{(6)}}\ep{z\a\mu}\ep{z\b\nu} \right],
    \end{aligned}
\end{equation}
where the coefficients $s^{(i)},\,i=1,\dots,6$ are
\begin{equation}
    \begin{aligned}
        &s^{(1)}=\frac{1}{9}\left( 3+4\cos\th+2\cos 2\th \right),\\
        &s^{(2)}=\frac{1}{3\sqrt{3}}\left( \cos\th-\cos 2\th \right),\quad s^{(3)}=-\frac{1}{3}\left( \sin\th+\sin 2\th \right),\\
        &s^{(4)}=-\frac{1}{\sqrt{3}}\left( 1-\cos\th \right)\sin\th,\\
        &s^{(5)}=\frac{1}{3}\left( 1-\cos\th \right)^2,\quad s^{(6)}=\sin^2\th.
    \end{aligned}
\end{equation}
From Eqs.~\eqref{MsToMsc_final}, \eqref{pq} and \eqref{L_eab}, we can derive the transform equation of the traceless part of the moment $M_{\a\b}^\pr$ as
\begin{equation}\label{MsToMsc_traceless}
    M_{\a\b}^{\pr sc}=\mathcal{L}_{\a\b\mu\nu}M_{\mu\nu}^{\pr s}.
\end{equation}
Again, we write this transformation under the representation of bases $\psi^{5\text{--}9}$ by using Eqs.~\eqref{BisTran}:
\begin{equation}\label{MsToMsc_traceless_Psi}
    \bm{M}^{\pr t+\h}=\bm{M}^{\pr sc}=\bm{\mathcal{L}}\cdot\bm{M}^{\pr s},
\end{equation}
where
\begin{widetext}
    \begin{equation}\label{L_psi}
        \bm{\mathcal{L}}=\ave{\frac{N-1}{N}}
        \left[ \begin{smallmatrix}
            \ave{\frac{1}{N-1}} + q\cos2\th & -q\sin2\th & 0 & 0 & 0 \\
            q\sin2\th &  \ave{\frac{1}{N-1}} + q\cos2\th & 0 & 0 & 0 \\
            0 & 0 & \ave{\frac{1}{N-1}} + q & 0 & 0 \\
            0 & 0 & 0 & \ave{\frac{1}{N-1}} +  q\cos\th & q\sin\th \\
            0 & 0 & 0 & -q\sin\th & \ave{\frac{1}{N-1}} + q\cos\th
        \end{smallmatrix} \right].
    \end{equation}
\end{widetext}

\emph{Stationary value of the velocity moment.}---The total transformation of $\bm{M}^{\pr}$ during $\left[ t,t+\h \right]$ can be derived by putting Eqs.~\eqref{MtToMs_Psi} and \eqref{MsToMsc_traceless_Psi} together:
\begin{equation}\label{MtToMsc_traceless_Psi}
    \bm{M}^{\pr t+\h}=\bm{\mathcal{L}}\cdot\left( \bm{M}^{\pr t}-2\T\h\dot{\bm{e}} \right).
\end{equation}
Setting $\bm{M}^{\pr t+\h}=\bm{M}^{\pr t}=\bm{M}^{\pr}$ in Eq.~\eqref{MtToMsc_traceless_Psi} yields a linear equation of the stationary value of the second-order velocity moment $\bm{M}^{\pr}$:
\begin{equation}
    \left( \bm{\mathcal{L}}-\bm{I}_{5\times5} \right)\cdot\bm{M}^{\pr}=2\T\h\bm{\mathcal{L}}\cdot\dot{\bm{e}}.
\end{equation}
Its solution is
\begin{equation}\label{SolMpr_Psi}
    \begin{aligned}
        &\bm{M}^{\pr}=2\T\h \bm{G}\cdot\dot{\bm{e}},\\
        &\bm{G}=
        \begin{bmatrix}
            \vp_e  & \vp_o &   0  &   0    &    0    \\ 
            -\vp_o & \vp_e &   0  &   0    &    0    \\ 
               0   &   0   & \psi_e &   0    &    0    \\ 
               0   &   0   &   0  & \phi_e & -\phi_o \\ 
               0   &   0   &   0  & \phi_o &  \phi_e    
        \end{bmatrix},
    \end{aligned}
\end{equation}
where
\begin{align*}
    &\vp_e=\ave{\frac{N}{N-1}}\frac{q\cos2\th-1}{\left( q\cos2\th-1 \right)^2+q^2\sin^22\th}+1,\\
    &\vp_o=\ave{\frac{N}{N-1}}\frac{q\sin2\th}{\left( q\cos2\th-1 \right)^2+q^2\sin^22\th},\\
    &\psi_e=\ave{\frac{N}{N-1}}\frac{1}{q-1}+1,\\
    &\phi_e=\ave{\frac{N}{N-1}}\frac{q\cos\th-1}{\left( q\cos\th-1 \right)^2+q^2\sin^2\th}+1,\\
    &\phi_o=\ave{\frac{N}{N-1}}\frac{q\sin\th}{\left( q\cos\th-1 \right)^2+q^2\sin^2\th}.
\end{align*}

\subsubsection{The kinetic viscosity}
Translating Eq.~\eqref{KinVisStress} into its form represented by $\psi^{5\text{--}9}$ and replacing the moment $\bm{M}^{\pr}$ by the solution Eq.~\eqref{SolMpr_Psi}, we obtain the kinetic viscous stress of the 3D-CSRD as:
\begin{equation}
    {\sigma^{\rm{v,kin}}}^I=2\rho\T\h\left( \frac{1}{2}\kru{IJ}-G^{IJ} \right)\dot{e}^J.
\end{equation}
Therefore the kinetic viscosity is
\begin{equation}\label{KinVisStress_Psi}
    {\eta^{\rm{kin}}}^{IJ}=2\rho\T\h\left( \frac{1}{2}\kru{IJ}-G^{IJ} \right),
\end{equation}
and can be expressed as the form of Eq.~\eqref{VisT_Psi}:
\begin{widetext}
    \begin{equation}\label{kinVis_M}
        \bm{\eta}^{\rm{kin}}=2
        \begin{bmatrix}
            \quad 0 \quad& \quad 0 \quad & \quad 0 \quad & \quad 0 \quad & \quad 0 \quad & \quad 0 \quad & \quad 0 \quad & \quad 0 \quad & \quad 0 \quad \\ 
            0 & 0 & 0 & 0 & 0 & 0 & 0 & 0 & 0 \\ 
            0 & 0 & 0 & 0 & 0 & 0 & 0 & 0 & 0 \\ 
            0 & 0 & 0 & 0 & 0 & 0 & 0 & 0 & 0 \\ 
            0 & 0 & 0 & 0 & \mu_1^{\rm{kin}} & \eta_1^{o,\rm{kin}} & 0 & 0 & 0 \\ 
            0 & 0 & 0 & 0 & -\eta_1^{o,\rm{kin}} & \mu_1^{\rm{kin}} & 0 & 0 & 0 \\ 
            0 & 0 & 0 & 0 & 0 & 0 & \mu_3^{\rm{kin}} & 0 & 0 \\ 
            0 & 0 & 0 & 0 & 0 & 0 & 0 & \mu_2^{\rm{kin}} & \eta_2^{o,\rm{kin}} \\ 
            0 & 0 & 0 & 0 & 0 & 0 & 0 & -\eta_2^{o,\rm{kin}} & \mu_2^{\rm{kin}} \\ 
        \end{bmatrix},
    \end{equation}
where the non-zero kinetic viscosities are as follows:
    \begin{equation}\label{kinVis}
        \begin{aligned}
                \mu_1^{\rm{kin}}&=\rho\T\h\left\{ \frac{\n\left( 1-q\cos2\th \right)}{\left( \n-1+e^{-\n} \right)\left[ \left( 1-q\cos2\th \right)^2+q^2\sin^22\th \right]} -\frac{1}{2}\right\},\\
                \mu_2^{\rm{kin}}&=\rho\T\h\left\{ \frac{\n\left( 1-q\cos\th \right)}{\left( \n-1+e^{-\n} \right)\left[ \left( 1-q\cos\th \right)^2+q^2\sin^2\th \right]} -\frac{1}{2}\right\},\\
                \mu_3^{\rm{kin}}&=\frac{1}{2}\rho\T\h\left[ \frac{5\n}{\left( \n-1+e^{-\n} \right)\left( 2-\cos\a-\cos2\a \right)} -1\right],\\
                \eta_1^{o,\rm{kin}}&=-\rho\T\h\frac{\n q\sin2\th}{\left( \n-1+e^{-\n} \right)\left[ \left( 1-q\cos2\th \right)^2+q^2\sin^22\th \right]},\\
                \eta_2^{o,\rm{kin}}&=\rho\T\h\frac{\n q\sin\th}{\left( \n-1+e^{-\n} \right)\left[ \left( 1-q\cos\th \right)^2+q^2\sin^2\th \right]}.
        \end{aligned}
    \end{equation}
\end{widetext}

\subsection{Navier-Stokes equation for the 3D-CSRD fluid}
The derivation of the mass continuity equation is straightforward. Substituting the mass flux Eq.~\eqref{jm_C} into the conservation equation of mass, we get the standard form of the continuity equation as
\begin{equation}\label{ContinutyEq}
    \pd{t}\rho+\pd{\a} \left( \rho u_\a\right)=0.
\end{equation}

Then we work on the conservation equation of momentum. According to Eqs.~\eqref{ColTab_C} and \eqref{KinTab_C}, the total momentum flux of CSRD is
\begin{equation}
    T_{\a\b}=\rho u_\a u_\b + P\kr{\a\b} - \sigma_{\a\b}^{\rm{v}},
\end{equation}
where $\sigma_{\a\b}^{\rm{v}}=\eta_{\a\b\mu\nu}\pd{\nu}u_\mu$. We can write down the following Navier-Stokes equation via the momentum flux, the conservation equation and the continuity equation:
\begin{equation}\label{NSEq_1}
    \rho\frac{\df u_\a}{\df t}=-\pd{\a}P+\pd{\b}\sigma_{\a\b}^{\rm{v}}.
\end{equation}

To obtain the standard form of Navier-Stokes equation for the 3D-CSRD, we have to transform the viscosity tensor back to the form expressed by the tensor product basis $\bm{e}_\a\otimes\bm{e}_\b$. The orthogonality of $\tau_{\a\b}^I$ (Eqs.~\eqref{Orth}) gives the relation between two expresses of the viscosity as: 
\begin{equation}\label{VisBackTran}
    \eta_{\a\b\mu\nu}=\frac{1}{2}\tau_{\a\b}^I\eta^{IJ}\tau_{\mu\nu}^J.
\end{equation} 
From Eqs.~\eqref{colVis_PisComp} and \eqref{kinVis}, we can derive $14$ non-zero viscosities of 3D-CSRD which are $\zeta$, $\eta_{R,1}$, $\eta_{R,2}$, $\mu_1$, $\mu_2$, $\mu_3$, $\eta_{s}^e$, $\eta_{Q,1}^e$, $\eta_{1}^o$, $\eta_{2}^o$, $\eta_{R}^o$, $\eta_{Q,2}^o$, $\eta_{Q,3}^o$, and $\eta_{A}^o$. We use Eq.~\eqref{VisBackTran} to transform the viscosity tensor part by part. For example, the $\mu_3$ part of viscosity tensor under basis $\left\{ \psi^I \right\}$ (i.e., $\mu_3\kru{I7}\kru{J7}$) is transformed back to $\mu_3\tau_{\a\b}^7\tau_{\mu\nu}^7$. For the sake of convenience, we expand the matrices $\tau_{\a\b}^{I}$ by Kronecker and Levi-Civita symbols. For the matrix $\tau_{\a\b}^7$, we have $\tau_{\a\b}^7=-\frac{1}{\sqrt{3}}\kr{\a\b}+\sqrt{3}\kr{z\a}\kr{z\b}$. Then the term of $\mu_3\tau_{\a\b}^7\tau_{\mu\nu}^7$ can be rearranged as 
\begin{equation}
    \begin{aligned}
        \mu_3\tau_{\a\b}^7\tau_{\mu\nu}^7=&\,\frac{1}{3}\mu_3\left( \kr{\a\b}\kr{\mu\nu}-3\kr{\a\b}\kr{z\mu}\kr{z\nu}\right.\\
        &\left.-3\kr{z\a}\kr{z\b}\kr{\mu\nu}+9\kr{z\a}\kr{z\b}\kr{z\mu}\kr{z\nu}\right),
    \end{aligned}
\end{equation}
After performing the similar procedure on other non-zero parts of the viscosity tensor, we get the viscosity tensor as
\begin{widetext}
    \begin{equation}
        \begin{aligned}
            \eta_{\a\b\mu\nu}=&\,\zeta\kr{\a\b}\kr{\mu\nu}+\mu_1\left( \delta^{\perp}_{\a\mu}\delta^{\perp}_{\b\nu}+\delta^{\perp}_{\a\nu}\delta^{\perp}_{\b\mu}-\delta^{\perp}_{\a\b}\delta^{\perp}_{\mu\nu} \right)+\mu_2\left( \delta^{\perp}_{\a\mu}\kr{z\b}\kr{z\nu}+\kr{z\a}\kr{z\mu}\delta^{\perp}_{\b\nu}+\delta^{\perp}_{\a\nu}\kr{z\b}\kr{z\mu}+\kr{z\a}\kr{z\nu}\delta^{\perp}_{\b\mu} \right)\\
            &+\mu_3\left( \frac{1}{3}\kr{\a\b}\kr{\mu\nu}-\kr{\a\b}\kr{z\mu}\kr{z\nu}-\kr{z\a}\kr{z\b}\kr{\mu\nu}+3\kr{z\a}\kr{z\b}\kr{z\mu}\kr{z\nu} \right)+\eta_{R,1}\left[ \delta_{\a\mu}\delta_{\b\nu}-\delta_{\a\nu}\delta_{\b\mu}-\left( \delta^{\perp}_{\a\mu}\delta^{\perp}_{\b\nu}-\delta^{\perp}_{\a\nu}\delta^{\perp}_{\b\mu} \right) \right]\\
            &+\eta_{R,2}\left( \delta^{\perp}_{\a\mu}\delta^{\perp}_{\b\nu}-\delta^{\perp}_{\a\nu}\delta^{\perp}_{\b\mu} \right)+\sqrt{2}\eta_s^e\left[ \frac{4}{3}\kr{\a\b}\kr{\mu\nu}-\left( \kr{\a\b}\delta_{\mu\nu}^{\perp}+\delta_{\a\b}^{\perp}\kr{\mu\nu} \right) \right]+2\eta_{Q,1}^e\left( \delta_{\a\mu}^{\perp}\kr{z\b}\kr{z\nu}-\kr{z\a}\kr{z\mu}\delta_{\b\nu}^{\perp} \right)\\
            &+\eta_1^o\left( \delta^{\perp}_{\a\mu}\ep{z\b\nu}+\delta^{\perp}_{\b\nu}\ep{z\a\mu} \right)-\eta_2^o\left( \kr{z\a}\kr{z\mu}\ep{z\b\nu}+\kr{z\a}\kr{z\nu}\ep{z\b\mu}+\kr{z\b}\kr{z\nu}\ep{z\a\mu}+\kr{z\b}\kr{z\mu}\ep{z\a\nu} \right)\\
            &+\eta_R^o\left( \ep{z\a\mu}\kr{\b\nu}+\kr{\a\mu}\ep{z\b\nu}-\ep{z\a\nu}\kr{\b\mu}-\kr{\a\nu}\ep{z\b\mu} \right)+2\eta_{Q,2}^o\left( \kr{\a\mu}\ep{z\b\nu}-\ep{z\a\mu}\kr{\b\nu} \right).
        \end{aligned}
    \end{equation}
Here, $\delta^{\perp}_{\a\b}$ is defined by $\delta^{\perp}_{\a\b}\triangleq\delta_{\a\b}-\kr{z\a}\kr{z\b}$ and the terms of $\eta_{Q,3}^o$ and $\eta_A^o$ have been absorbed into $\eta_{Q,2}^o$ term by using the relations $\eta_{Q,3}^o=2\eta_{Q,2}^o/\sqrt{3}$ and $\eta_A^o=-\frac{4}{\sqrt{6}}\eta_{Q,2}^o$ (see Eqs.~\eqref{colVis_PisComp}). Proceeding, the viscous stress constitutive relation, $\sigma_{\a\b}^{\rm{v}}=\eta_{\a\b\mu\nu}\pd{\nu}u_\mu$, of the 3D-CSRD is written as,
    \begin{equation}
        \begin{aligned}
            \sigma_{\a\b}^{\rm{v}}=&\,\zeta\bm{\nabla}\cdot\bm{u}+\mu_1\left( \pdv{\b}u_\a^\perp+\pdv{\a}u_\b^\perp-\krv{\a\b}\bm{\nabla}_\perp\cdot\bm{u}^\perp \right)+\mu_2\left[ \kr{z\b}\left( \pd{z}u_\a^\perp+\pd{\a}u_z^\perp \right)+\kr{z\a}\left( \pd{z}u_\b^\perp+\pd{\b}u_z^\perp \right) \right]\\
                &+\mu_3\left( \krv{\a\b}\bm{\nabla}_\perp\cdot\bm{u}^\perp+2\kr{z\a}\kr{z\b}\pd{z}u_z-\frac{2}{3}\kr{\a\b}\bm{\nabla}\cdot\bm{u} \right)+\eta_{R,1}\left[ \pd{\b}u_\a-\pd{\a}u_\b-\left( \pdv{\b}u_\a^\perp-\pdv{\a}u_\b^\perp \right) \right]\\
                &+\eta_{R,2}\left( \pdv{\b}u_\a^\perp-\pdv{\a}u_\b^\perp \right)+\sqrt{2}\eta_s^e\left[ \frac{4}{3}\kr{\a\b}-\left( \kr{\a\b}\bm{\nabla}_\perp\cdot\bm{u}^\perp+\krv{\a\b}\bm{\nabla}\cdot\bm{u} \right) \right]+2\eta_{Q,1}^e\left( \kr{z\b}\pd{z}u_\a^\perp-\kr{z\a}\pdv{\b}u_z \right)\\
                &+\eta_1^o\left( \pdu{\b}{\ast}u_\a^\perp +\pdv{\b}u_\a^\ast\right)-\eta_2^o\left[ \kr{z\a}\left( \pdu{\b}{\ast}u_z+\pd{z}u_\b^\ast \right)+\kr{z\b}\left( \pdu{\a}{\ast}u_z+\pd{z}u_\a^\ast \right) \right]\\
                &+\eta_R^o\left( \pd{\b}u_\a^\ast+\pdu{\b}{\ast}u_\a-\pd{\a}u_\b^\ast+\pdu{\a}{\ast}u_\b \right)+2\eta_{Q,2}^o\left( \pdu{\b}{\ast}u_\a-\pd{\b}u_\a^\ast\right).
        \end{aligned}
    \end{equation}
\end{widetext}
Here, we have defined the notations $u_\a^\ast\triangleq\ep{z\a\b}u_\b$ and $u_\a^{\perp}\triangleq \krv{\a\b}u_\b$ (these notations are also applied to $\bm{\nabla}=\left( \pd{x},\pd{y},\pd{z} \right)^\top$ in the same way). Finally, with this constitutive relation, the Navier-Stokes equation for 3D-CSRD becomes
\begin{equation}\label{NSEq}
    \begin{aligned}
        \rho\frac{\mathrm{d}\bm{u}}{\mathrm{d}t}=&\,-\bm{\nabla} P+\hat{\eta}_b\bm{\nabla}\left( \bm{\nabla}\cdot\bm{u} \right)+\hat{\eta}_{zb}\left( \bm{\nabla}\pd{z}u_z+\hat{\bm{e}}_z\pd{z}\bm{\nabla}\cdot\bm{u}\right)\\
        &+\hat{\eta}\bm{\nabla}^2\bm{u}+\hat{\eta}_{zs1}\hat{\bm{e}}_z\bm{\nabla}^2u_z+\hat{\eta}_{zs2}\pdu{z}{2}\bm{u}+\hat{\eta}_{zs3}\hat{\bm{e}}_z\pdu{z}{2}u_z\\
        &+\hat{\eta}_{o}\bm{\nabla}^2\bm{u}^\ast+\hat{\eta}_{ob}\left[ \bm{\nabla}\left( \bm{\nabla}\cdot\bm{u}^\ast \right)+\bm{\nabla}^\ast\left( \bm{\nabla}\cdot\bm{u} \right) \right]\\
        &+\hat{\eta}_{zo}\left( \pdu{z}{2}\bm{u}^\ast+\hat{\bm{e}}_z\pd{z}\bm{\nabla}\cdot\bm{u}^\ast+\bm{\nabla}^\ast\pd{z}u_z \right),
    \end{aligned}
\end{equation}
where the following coefficients are introduced:
\begin{align*}
    \hat{\eta}&\triangleq\mu_1+\eta_{R,2}=\mu_1^{\rm{kin}}+\eta_1^{\rm{col}}-\frac{1}{\sqrt{3}}\eta_2^{\rm{col}},\\
    \hat{\eta}_b&\triangleq\zeta+\frac{1}{3}\mu_3-\eta_{R,2}-\frac{2\sqrt{2}}{3}\eta_s^e=\frac{1}{3}\mu_3^{\rm{kin}},\\
    \hat{\eta}_{zs1}&\triangleq\mu_2-\mu_1+\eta_{R,1}-\eta_{R,2}-2\eta_{Q,1}^e\\
                    &=\mu_2^{\rm{kin}}-\mu_1^{\rm{kin}}+\frac{\sqrt{3}}{2}\eta_2^{\rm{col}},\\
    \hat{\eta}_{zs2}&\triangleq\mu_2-\mu_1+\eta_{R,1}-\eta_{R,2}+2\eta_{Q,1}^e\\
                    &=\mu_2^{\rm{kin}}-\mu_1^{\rm{kin}}-\frac{\sqrt{3}}{2}\eta_2^{\rm{col}},\\
    \hat{\eta}_{zs3}&\triangleq2\mu_1+2\mu_3-4\mu_2+\eta_{R,2}-\eta_{R,1}\\
                    &=2\left( \mu_1^{\rm{kin}}+\mu_3^{\rm{kin}}-2\mu_2^{\rm{kin}} \right)-\frac{7\sqrt{3}}{12}\eta_2^{\rm{col}},\\
    \hat{\eta}_{zb}&\triangleq\mu_2-\mu_3+\eta_{R,2}-\eta_{R,1}+\sqrt{2}\eta_{s}^e\\
                    &=\mu_2^{\rm{kin}}-\mu_3^{\rm{kin}},\\
    \hat{\eta}_o&\triangleq\frac{1}{2}\eta_1^o+\eta_R^o-2\eta_{Q,2}^o=\frac{1}{2}\eta_1^{o,\rm{kin}}+\eta_3^{\rm{col}},\\
    \hat{\eta}_{ob}&\triangleq\frac{1}{2}\eta_1^o-\eta_R^o=\frac{1}{2}\eta_1^{o,\rm{kin}},\\
    \hat{\eta}_{zo}&\triangleq-\left( \frac{1}{2}\eta_1^o+\eta_2^o \right)=-\left( \frac{1}{2}\eta_1^{o,\rm{kin}}+\eta_2^{o,\rm{kin}} \right).
\end{align*}

\subsubsection{Simplified Navier-Stokes equation}
The Navier-Stokes equation will be significantly simplified if we choose a small additional rotation angle $\theta$ and set $\omega=2\pi/3$. Under the condition of $\omega=2\pi/3$, $\eta_2^{\rm{col}}$ and $\eta_3^{\rm{col}}$ become zero (see Eqs.~\eqref{colVis_origin}) such that in Eqs.~\eqref{colVis_PisComp} $\eta_s^{e,\rm{col}}$, $\eta_{Q,1}^{e,\rm{col}}$, and all collisional odd viscosities vanish; while the equality that $\mu_1^{\rm{col}}=\mu_2^{\rm{col}}=\mu_3^{\rm{col}}=\eta_{R,1}^{\rm{col}}=\eta_{R,2}^{\rm{col}}$ holds. In Eqs.~\eqref{kinVis}, the small $\theta$ limit yields two approximate relations for the kinetic viscosities $\mu_1^{\rm{kin}}=\mu_2^{\rm{kin}}=\mu_3^{\rm{kin}}$ and $\eta_1^{o,\rm{kin}}=-2\eta_2^{o,\rm{kin}}$. By defining $\mu\triangleq\mu_1=\mu_2=\mu_3$, $\eta_R\triangleq\eta_{R,1}=\eta_{R,2}$, and $\eta_o\triangleq\eta_1^{o}=-2\eta_2^{o}$, the viscosity tensor now reads
\begin{equation}
    \begin{aligned}
        &\eta_{\a\b\mu\nu}\\
        =&\,\zeta\kr{\a\b}\kr{\mu\nu}+\eta_R\left( \kr{\a\mu}\kr{\b\nu}-\kr{\a\nu}\kr{\b\mu} \right)\\
        &+\mu\left( \kr{\a\mu}\kr{\b\nu}+\kr{\a\nu}\kr{\b\mu}-\frac{2}{3}\kr{\a\b}\kr{\mu\nu} \right)\\
        &+\frac{1}{2}\eta_o\left( \ep{z\a\mu}\kr{\b\nu}+\ep{z\a\nu}\kr{\b\mu}+\ep{z\b\mu}\kr{\a\nu}+\ep{z\b\nu}\kr{\a\mu} \right).
    \end{aligned}
\end{equation}
The Navier-Stokes equation is simplified as
\begin{equation}
    \begin{aligned}
        \rho\frac{\mathrm{d}\bm{u}}{\mathrm{d}t}=&\,-\bm{\nabla} p+\hat{\eta}\bm{\nabla}^2\bm{u}+\hat{\eta}_b\bm{\nabla}\left( \bm{\nabla}\cdot\bm{u} \right)\\
        &+\hat{\eta}_o\left[ \bm{\nabla}^2\bm{u}^\ast+\bm{\nabla}\left( \bm{\nabla}\cdot\bm{u}^\ast \right)+\bm{\nabla}^\ast\left( \bm{\nabla}\cdot\bm{u} \right) \right],
    \end{aligned}
\end{equation}
where $\hat{\eta}=\eta_1^{\rm{col}}+\mu^{\rm{kin}},\,\hat{\eta}_b=\frac{1}{3}\mu^{\rm{kin}},\,$ and $\hat{\eta}_o=\frac{1}{2}\eta_1^{o,\rm{kin}}$.

\section{Viscosities of 3D-CSRD fluid}\label{Appendix}
We summarize the theoretical expressions and the numerical results of all the viscosities for 3D-CSRD fluid in Table.~\ref{VisTable}. For simpleness, the quantities $q$, $\eta_1^{\rm{col}}$, $\eta_2^{\rm{col}}$, $\eta_3^{\rm{col}}$ are used in the table and their expressions are:
\begin{align*}
    &q=\frac{1}{5}\left( 1+2\cos\omega+2\cos 2\omega \right),\\
    &\eta_1^{\rm{col}}=\frac{m\left( \n-1+e^{-\n} \right)}{108l\h}\left[ 9 - \left( 1+2\cos\omega \right)\left( 1+2\cos\theta \right) \right],\\
    &\eta_2^{\rm{col}}=-\frac{\sqrt{3}m\left( \n-1+e^{-\n} \right)}{108l\h}\left( 1+2\cos\omega \right)\left( 1-\cos\theta \right),\\
    &\eta_3^{\rm{col}}=\frac{m\left( \n-1+e^{-\n} \right)}{36l\h}\left( 1+2\cos\omega \right)\sin\theta.
\end{align*}

\qquad\newline
\qquad\newline
\begin{widetext}
\begin{longtable*}{ccccc}
    \caption{Viscosities of 3D-CSRD fluid. {  Parameters: $\h=0.1$, $\lambda=10$, $\omega=\pi/3$, $k_BT=1$}.}\label{VisTable}\\
    \hline\hline
    Viscosity & Even/Odd & Kinetic Part & Collisional Part & Simulation Result \\ \hline
    \\
    $\zeta$   &   Even   &     $0$      & $\frac{1}{3}\eta_1^{\rm{col}}$ & \adjustbox{valign=c}{\includegraphics[width=0.3\textwidth,keepaspectratio]{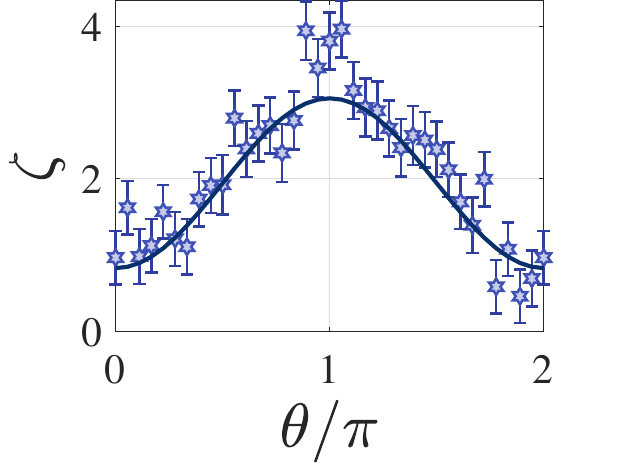}}\\
    $\eta_{R,1}$ & Even  &     $0$      & $\frac{1}{2}\eta_1^{\rm{col}}+\frac{1}{4\sqrt{3}}\eta_2^{\rm{col}}$ & \adjustbox{valign=c}{\includegraphics[width=0.3\textwidth,keepaspectratio]{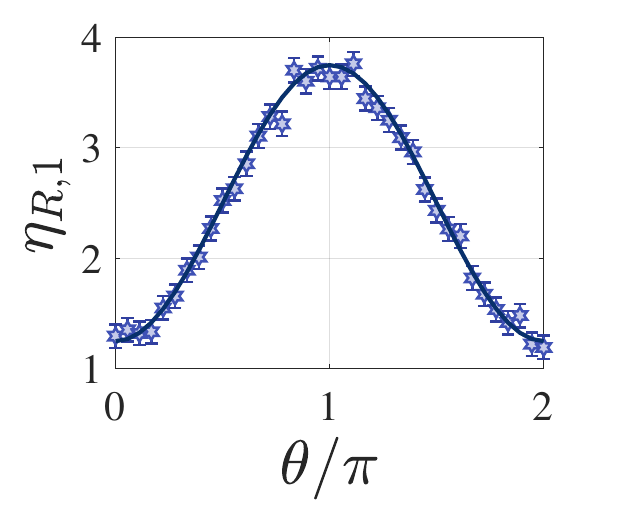}}\\
    $\eta_{R,2}$ & Even  &     $0$      & $\frac{1}{2}\eta_1^{\rm{col}}-\frac{1}{2\sqrt{3}}\eta_2^{\rm{col}}$ & \adjustbox{valign=c}{\includegraphics[width=0.3\textwidth,keepaspectratio]{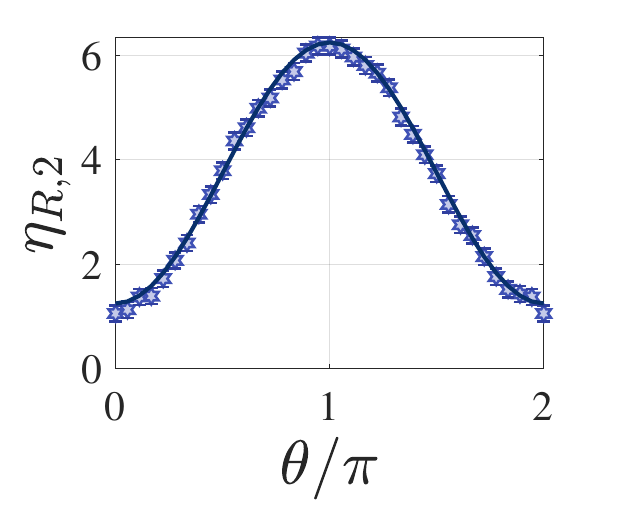}}\\
    $\mu_1$   &   Even   &     $\frac{\rho k_BT\h}{m}\left\{ \frac{\n\left( 1-q\cos2\th \right)}{\left( \n-1+e^{-\n} \right)\left[ \left( 1-q\cos2\th \right)^2+q^2\sin^22\th \right]} -\frac{1}{2}\right\}$ & $\frac{1}{2}\eta_1^{\rm{col}}-\frac{1}{2\sqrt{3}}\eta_2^{\rm{col}}$ & \adjustbox{valign=c}{\includegraphics[width=0.3\textwidth,keepaspectratio]{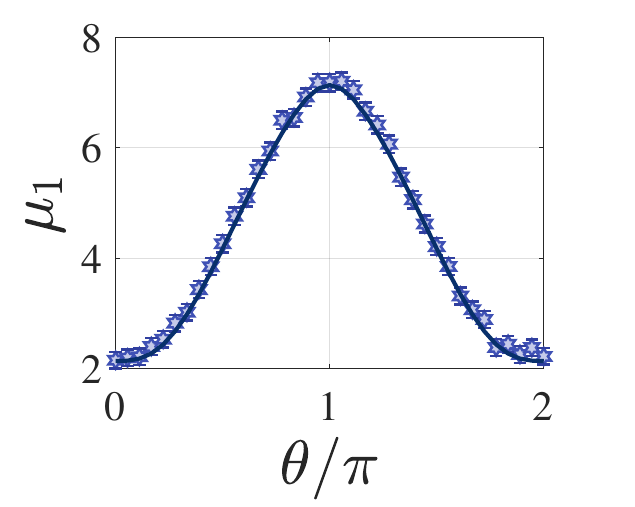}}\\
    $\mu_2$   &   Even   &     $\frac{\rho k_BT\h}{m}\left\{ \frac{\n\left( 1-q\cos\th \right)}{\left( \n-1+e^{-\n} \right)\left[ \left( 1-q\cos\th \right)^2+q^2\sin^2\th \right]} -\frac{1}{2}\right\}$ & $\frac{1}{2}\eta_1^{\rm{col}}+\frac{1}{4\sqrt{3}}\eta_2^{\rm{col}}$ & \adjustbox{valign=c}{\includegraphics[width=0.3\textwidth,keepaspectratio]{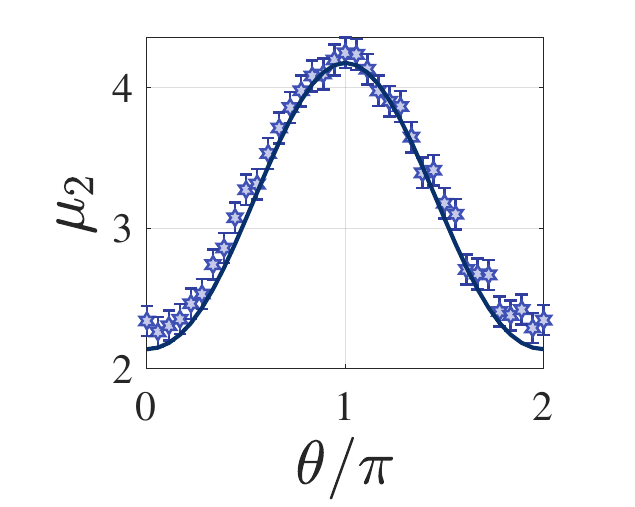}}\\
    $\mu_3$   &   Even   &     $\frac{\rho k_BT\h}{2m}\left[ \frac{5\n}{\left( \n-1+e^{-\n} \right)\left( 2-\cos\omega-\cos2\omega \right)} -1\right]$ & $\frac{1}{2}\eta_1^{\rm{col}}+\frac{1}{2\sqrt{3}}\eta_2^{\rm{col}}$ & \adjustbox{valign=c}{\includegraphics[width=0.3\textwidth,keepaspectratio]{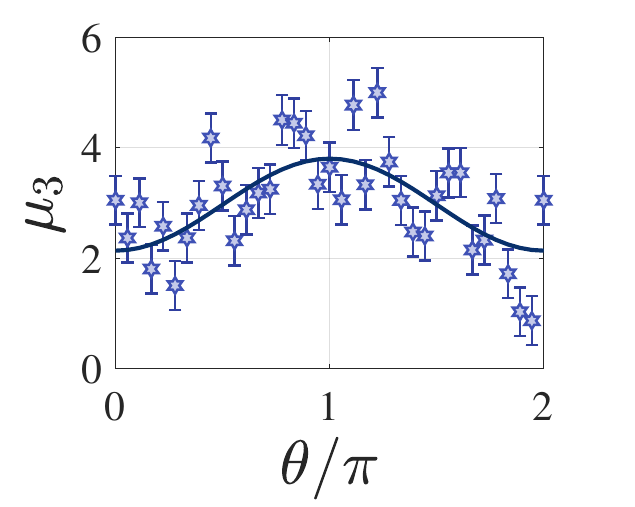}}\\
    $\eta_A^e$ & Even  &     $0$      & $0$ & \adjustbox{valign=c}{\includegraphics[width=0.3\textwidth,keepaspectratio]{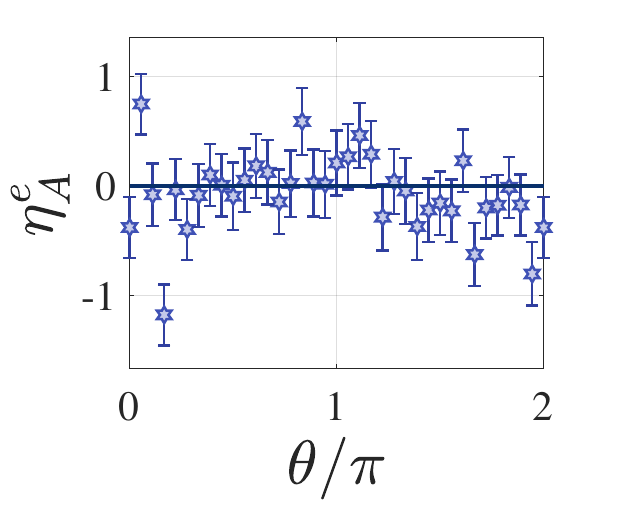}}\\
    $\eta_{s}^e$ & Even  &     $0$      & $\frac{1}{\sqrt{6}}\eta_2^{\rm{col}}$ & \adjustbox{valign=c}{\includegraphics[width=0.3\textwidth,keepaspectratio]{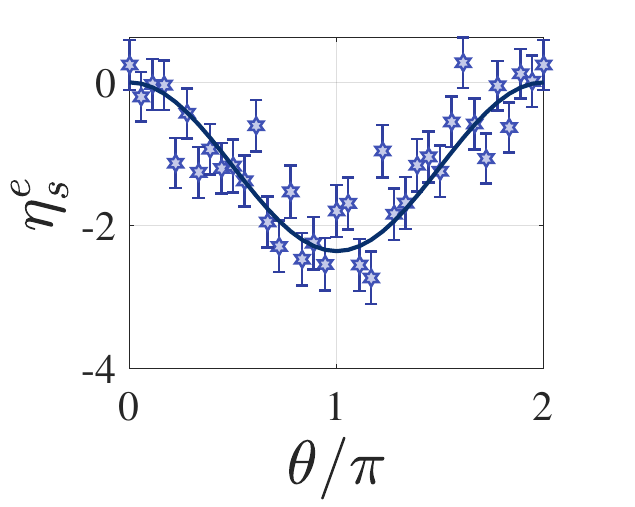}}\\
    $\eta_{Q,1}^e$ & Even  &     $0$      & $-\frac{3}{4\sqrt{3}}\eta_2^{\rm{col}}$ & \adjustbox{valign=c}{\includegraphics[width=0.3\textwidth,keepaspectratio]{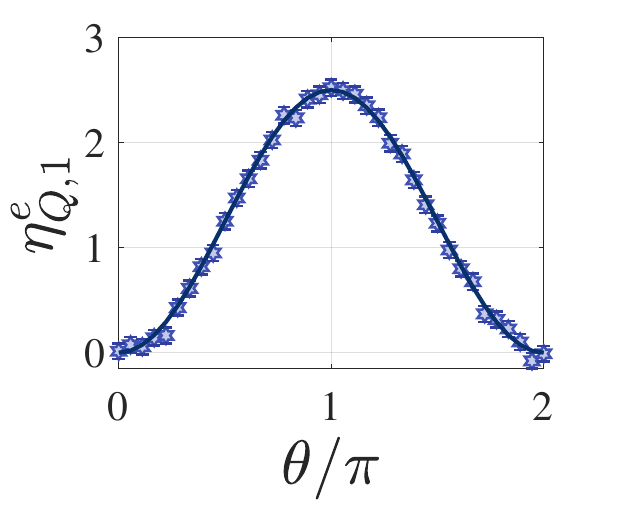}}\\
    $\eta_{Q,2}^e$ & Even  &     $0$      & $0$ & \adjustbox{valign=c}{\includegraphics[width=0.3\textwidth,keepaspectratio]{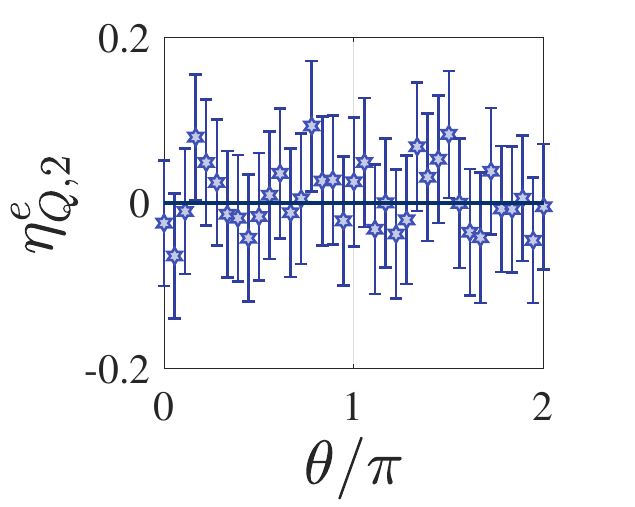}}\\
    $\eta_{Q,3}^e$ & Even  &     $0$      & $0$ & \adjustbox{valign=c}{\includegraphics[width=0.3\textwidth,keepaspectratio]{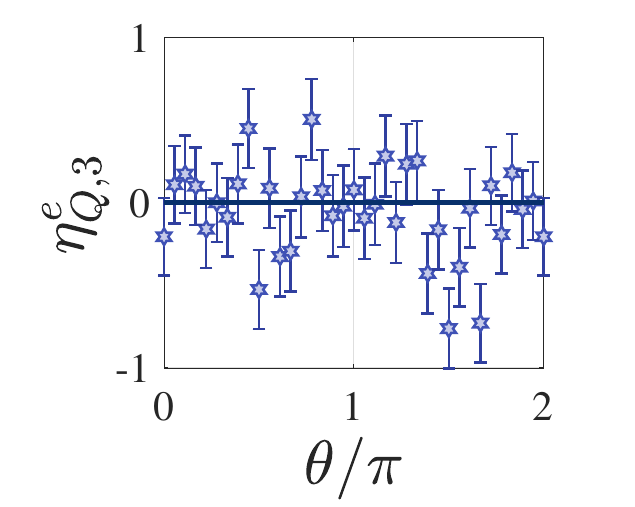}}\\
    $\eta_1^o$     & Odd   & $-\frac{\rho k_BT\h\n q\sin2\th}{m\left( \n-1+e^{-\n} \right)\left[ \left( 1-q\cos2\th \right)^2+q^2\sin^22\th \right]}$ & $\frac{1}{2}\eta_3^{\rm{col}}$ & \adjustbox{valign=c}{\includegraphics[width=0.3\textwidth,keepaspectratio]{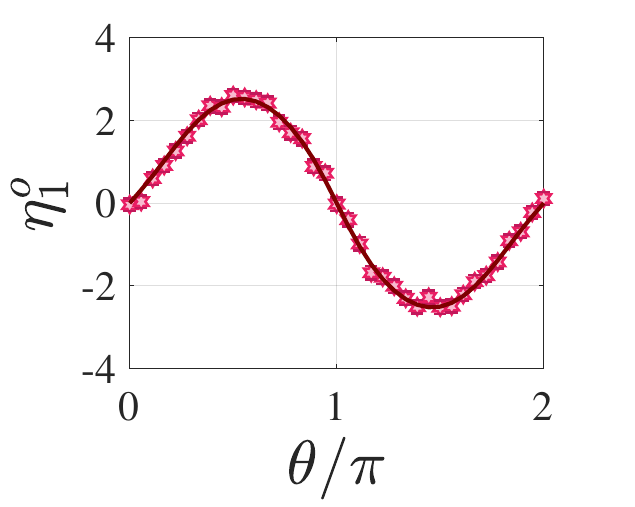}}\\
    $\eta_2^o$     & Odd   & $\frac{\rho k_BT\h\n q\sin\th}{m\left( \n-1+e^{-\n} \right)\left[ \left( 1-q\cos\th \right)^2+q^2\sin^2\th \right]}$ & $-\frac{1}{4}\eta_3^{\rm{col}}$ & \adjustbox{valign=c}{\includegraphics[width=0.3\textwidth,keepaspectratio]{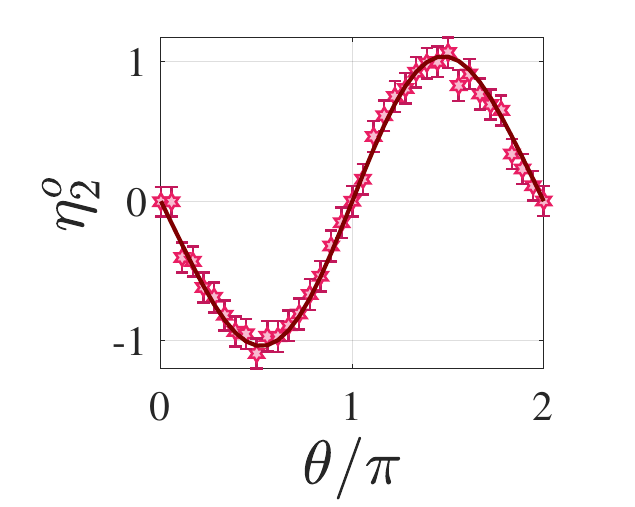}}\\
    $\eta_R^o$     & Odd   & $0$ & $\frac{1}{4}\eta_3^{\rm{col}}$ & \adjustbox{valign=c}{\includegraphics[width=0.3\textwidth,keepaspectratio]{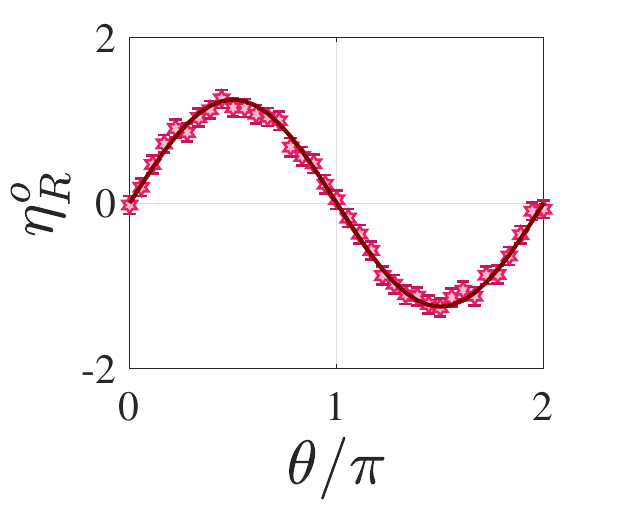}}\\
    $\eta_A^o$     & Odd   & $0$ & $\frac{1}{\sqrt{6}}\eta_3^{\rm{col}}$ & \adjustbox{valign=c}{\includegraphics[width=0.3\textwidth,keepaspectratio]{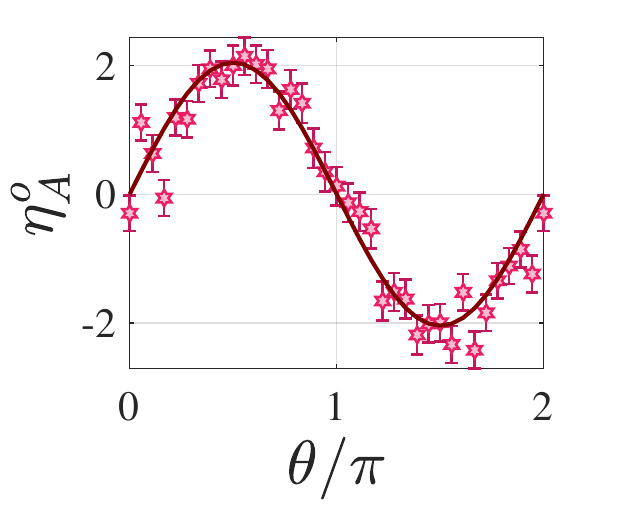}}\\
    $\eta_s^o$     & Odd   & $0$ & $0$ & \adjustbox{valign=c}{\includegraphics[width=0.3\textwidth,keepaspectratio]{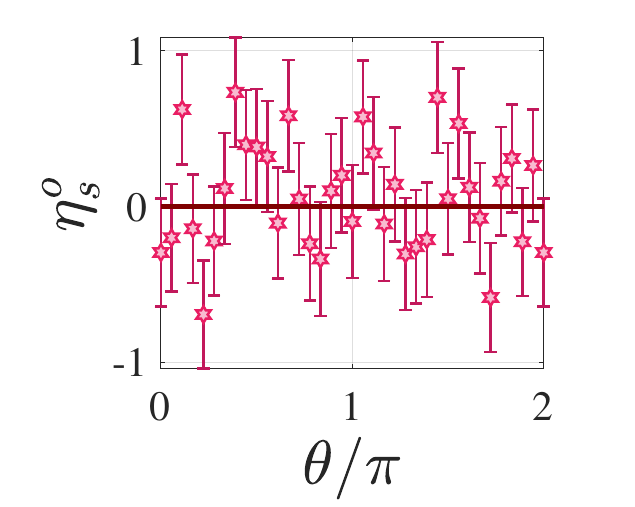}}\\
    $\eta_{Q,1}^o$     & Odd   & $0$ & $0$ & \adjustbox{valign=c}{\includegraphics[width=0.3\textwidth,keepaspectratio]{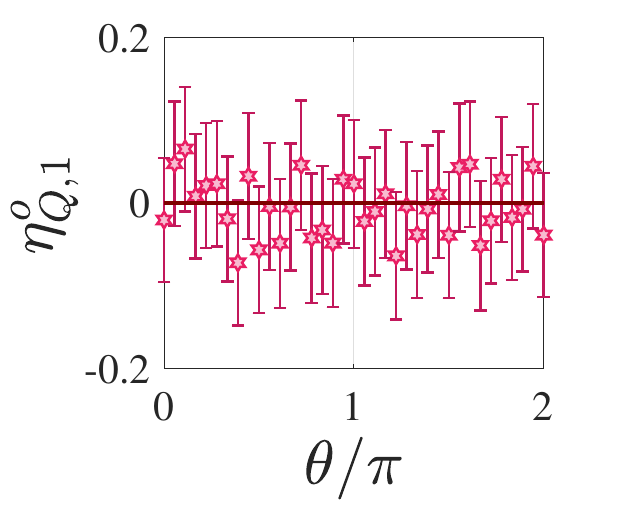}}\\
    $\eta_{Q,2}^o$     & Odd   & $0$ & $-\frac{1}{4}\eta_3^{\rm{col}}$ & \adjustbox{valign=c}{\includegraphics[width=0.3\textwidth,keepaspectratio]{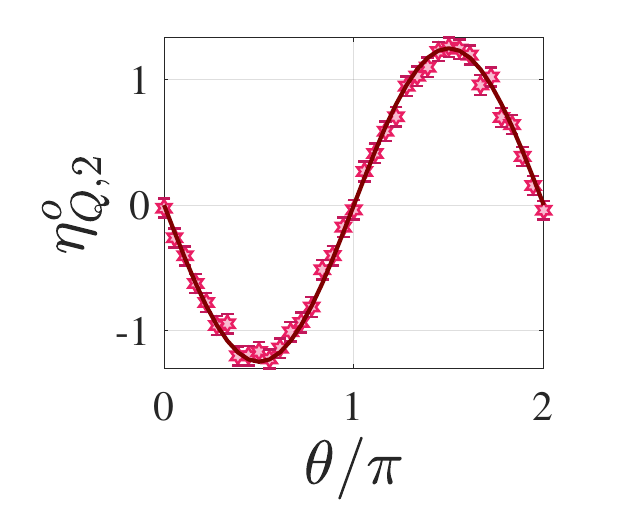}}\\
    $\eta_{Q,3}^o$     & Odd   & $0$ & $-\frac{1}{2\sqrt{3}}\eta_3^{\rm{col}}$ & \adjustbox{valign=c}{\includegraphics[width=0.3\textwidth,keepaspectratio]{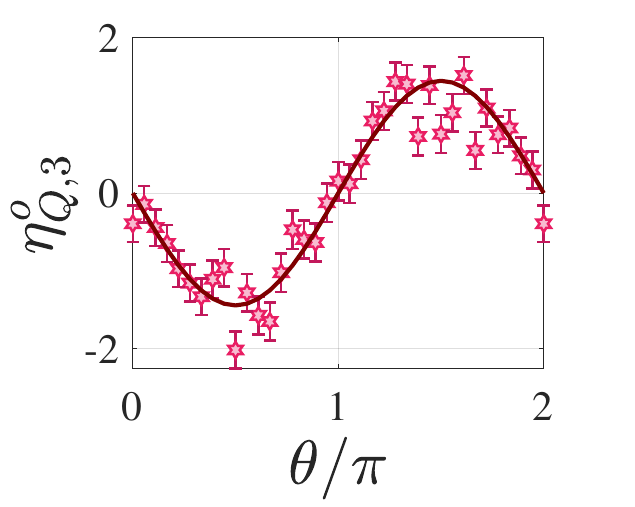}}\\
    \hline\hline
\end{longtable*}
Here, for the simulation result of bulk viscosity $\zeta$, we only disply its collisional part because our current method cannot measure a kinetic bulk viscosity with a sufficient precision.
\end{widetext}


\end{document}